\documentclass{aa}
\usepackage{amsmath,amssymb,bm,booktabs,graphicx,txfonts}
\usepackage{hyperref}
\providecommand{\aap}{A\&A}
\providecommand{\apj}{ApJ}
\providecommand{\apjl}{ApJL}
\providecommand{\aj}{AJ}
\providecommand{\mnras}{MNRAS}
\providecommand{\prd}{Phys.~Rev.~D}
\providecommand{\pasp}{PASP}
\providecommand{\araa}{ARA\&A}
\newcommand{\kms}{\ensuremath{\mathrm{km\,s^{-1}}}}

\newcommand{\pantheon}{Pantheon$+$}

\hypersetup{colorlinks=true,linkcolor=blue,citecolor=blue,filecolor=blue,urlcolor=blue,
pdftitle={Anisotropy and tracer-profile assumptions in aperture-based strong-lens dynamics},
pdfauthor={Jian Hu, Yi Liu, Chen Wang, Jian-Ping Hu, Jia Zhao, Liangqing Zhu, Zhongmu Li}}
\makeatletter
\let\revision@enddocument\enddocument
\def\enddocument{\NAT@swatrue\let\bibcite\NAT@testdef\revision@enddocument}
\renewcommand*\aa@pageof{, page \thepage{} of \pageref*{LastPage}}
\makeatother
\begin{document}

\title{Anisotropy and tracer-profile assumptions in aperture-based strong-lens dynamics}
\subtitle{Conditional inference from supernova distance ratios and unresolved stellar kinematics}
\titlerunning{Anisotropy and tracer-profile sensitivity}
\authorrunning{Hu et al.}
\author{Jian Hu\inst{1,\ast}\and Yi Liu\inst{2}\and Chen Wang\inst{3,\ast}\and Jian-Ping Hu\inst{4,5}\and Jia Zhao\inst{6}\and Liangqing Zhu\inst{7}\and Zhongmu Li\inst{1}}
\institute{
Institute of Astronomy and Information, Dali University, Dali 671003, People's Republic of China\\\email{dg1626002@smail.nju.edu.cn}\and Department of Engineering, Dali University, Dali 671003, People's Republic of China\and School of General Education and Foreign Languages, Anhui Institute of Information Technology, Wuhu 241100, People's Republic of China\\\email{wangchen@pmo.ac.cn}\and Ministry of Education Key Laboratory for Nonequilibrium Synthesis and Modulation of Condensed Matter, School of Physics, Xi'an Jiaotong University, Xianning West Road, Xi'an, People's Republic of China\and Key Laboratory of Modern Astronomy and Astrophysics (Nanjing University), Ministry of Education, Xianlin Road, Nanjing, People's Republic of China\and Water Affairs Bureau of Dali Bai Autonomous Prefecture, Dali 671003, People's Republic of China\and State Grid Zhejiang Electric Power Co., Ltd. Lishui Power Supply Company, Lishui 323050, People's Republic of China\\[2pt]\textit{$\ast$ Corresponding authors: Jian Hu and Chen Wang.}}
\date{Received ---; accepted ---}

\abstract
{Single-aperture stellar kinematics jointly constrain the mass distribution, tracer light profile, and velocity anisotropy. Their coupling affects the interpretation of an inferred anisotropy.}
{We quantify how tracer-profile assumptions and data treatment affect common-anisotropy comparisons and inferred mass-density slopes in a compiled galaxy-scale lens sample.}
{We analyse 107 lenses using distinct \pantheon\ supernova counterparts, spatial flatness, distance duality, and the full supernova covariance. We compare the individual-slope power-law tracer with Hernquist and S\'ersic profiles using measured effective radii. Numerical Jeans solutions, likelihood comparisons, posterior distributions, and controlled recovery experiments quantify the effects of tracer representation.}
{The power-law fixed-minus-free Bayesian information criterion differences are 6.74 for isotropy and 41.51 for fixed $\beta=0.18$, compared with $-3.89$ and $-2.26$ for Hernquist and $-4.67$ and $-4.30$ for S\'ersic $n=4$. The preference also changes with S\'ersic index. Removing 12 systems with linearised distance-ratio uncertainty above 20\% reduces the power-law isotropy comparison to 2.51. In isotropic S\'ersic $n=4$ and Hernquist simulations, power-law fits yield $\Delta\mathrm{BIC}_0>6$ in 60.0\% and 50.0\% of realisations, respectively, compared with 0.5\% and 1.0\% for matched-tracer fits.}
{The preference for a common free anisotropy parameter depends on the tracer representation and sample composition. Controlled recovery identifies tracer mismatch as one mechanism that can generate this preference within the adopted aperture models. Physical interpretation requires joint treatment of photometry, distance uncertainties, and kinematic weighting.}
\maketitle

\section{Introduction}\label{sec:intro}

Strong gravitational lensing (SGL) connects the projected mass of a foreground galaxy to the geometry between the observer, lens, and background source. Combining lensing observables with stellar kinematics constrains the mass distribution of early-type galaxies (ETGs) and provides information on cosmological distances \citep{Treu2010,Koopmans2006,Cao2015}. Time-delay cosmography additionally uses measured delays between variable images to infer distances and the Hubble constant $H_0$ \citep{Suyu2017,Wong2020}. Here we study galaxy--galaxy lenses using supernova (SN) distance ratios and aperture-integrated stellar dispersions.

Distance-ratio applications of galaxy-scale lenses have constrained dark-energy models using joint lensing and dynamical observables \citep{Biesiada2006,Biesiada2010,Cao2012}. Time-delay analyses offer a complementary route to the distance scale, with the result depending on the mass model and the treatment of external convergence \citep{Birrer2016,Millon2020}. Independent distance probes are of particular interest given the difference between cosmic-microwave-background and local distance-ladder determinations \citep{Planck2020,Riess2022}. Reviews discuss both observational systematics and possible extensions of the cosmological model \citep{DiValentino2021,Abdalla2022,Perivolaropoulos2022}.

A central difficulty is that lensing and dynamical observables constrain different combinations of the mass profile and stellar orbital structure. Imaging alone admits the mass-sheet degeneracy, while dynamical inference depends on the tracer light distribution, projection, and velocity anisotropy \citep{Falco1985,Schneider2013,Birrer2020}. Increasing the sample size reduces statistical errors but does not automatically control errors from an inadequate model. Potential biases must therefore be identified, constrained with additional data, and included in the uncertainty budget. Stellar anisotropy is not intrinsically a systematic error: the relevant issue is whether the assumed anisotropy model and its uncertainty are adequate for the data.

The power-law description is an effective representation of a composite stellar and dark-matter mass distribution \citep{Treu2002,Navarro1997}. Lensing and dynamical studies find approximately isothermal total profiles over the probed central regions, together with variations among galaxies \citep{Koopmans2009,Thomas2009,Tortora2014}. Accretion and mergers can change galaxy sizes and the radial weighting of the stellar component \citep{Naab2009,Oser2010}. Analyses of lens samples have therefore investigated the dependence of the fitted density slope on redshift, stellar mass, size, and surface density \citep{Ruff2011,Bolton2012,Sonnenfeld2013}; recent measurements extend such structural comparisons to lensed-quasar deflectors \citep{Guo2026}. These results motivate keeping tracer structure, total density slope, and orbital anisotropy distinct in a compiled aperture analysis.

Current analyses do not generally fix all lenses to one universal anisotropy value. The TDCOSMO collaboration's 2025 analysis uses a population treatment of spatially constant anisotropy, with uncertainty in the population mean and scatter \citep{TDCOSMO2025}. Spatial constancy within a galaxy is distinct from fixing the same known value across galaxies. \citet{Fores2026} examine anisotropy alongside point-spread-function (PSF), light-profile, and kinematic-measurement effects. Using simulated galaxies and spatially resolved kinematics, \citet{Verma2026} find that the recovery of $H_0$ depends on the adopted anisotropy family and on whether lensing and dynamics are modelled jointly. These studies emphasise the interaction between anisotropy and other modelling choices, rather than the inevitability of an anisotropy-induced bias.

Resolved spectroscopy is also advancing rapidly. The James Webb Space Telescope (JWST) Near-Infrared Spectrograph (NIRSpec) integral field unit (IFU) has provided two-dimensional stellar kinematics for RXJ1131$-$1231 \citep{Shajib2026}. \citet{Knabel2026} extend such measurements to HE0435$-$1223, PG1115$+$080, and WFI2033$-$4723, while \citet{Mozumdar2026} present resolved measurements for B1608$+$656 and SDSSJ1206$+$4332, including rotation and neighbouring deflectors. These observations supply radial and spatial information that a single aperture dispersion cannot provide. Our analysis quantifies modelling sensitivity in the complementary regime of compiled aperture-integrated measurements.

The survey designs for Rubin/LSST, Euclid, and the China Space Station Telescope (CSST) anticipate much larger imaging samples \citep{Ivezic2019,Laureijs2011,Gong2019}. Lens-population and cosmological forecasts illustrate the resulting statistical opportunity \citep{Collett2015,Chen2025forecast}. Hierarchical analyses explicitly combine population-level lens structure with cosmological inference \citep{stad3514,Geng2026}. This prospect also motivates quantitative checks of how aperture modelling assumptions affect a common-anisotropy comparison.

We use the individual photometric slopes and lens compilation of \citet{stz1902}, together with supernova-based distance ratios \citep{Hu2023p,Hu2023,Geng2025}, to quantify the sensitivity of a common-anisotropy preference to the tracer profile and compiled-data approximations. Isotropy and $\beta=0.18$ are reference models; the historical $0.18\pm0.13$ prescription \citep{Gerhard2001,stz1902} is not a universal description of contemporary lens analyses. Our comparison also separates a fixed value from a probability distribution over values.

Recent work shows how anisotropy interacts with light-profile modelling and kinematic weighting \citep{Fores2026,Verma2026}. We quantify these interactions in a different observational regime: a catalogue of unresolved aperture dispersions with supernova-derived distance inputs. We compare tracer families on the same matched catalogue and covariance, measure the resulting changes in anisotropy and density-slope inference, and use controlled recovery experiments to assess whether tracer mismatch can generate an effective anisotropy preference. The photometric information entering each tracer family is specified in Sect.~\ref{sec:tracers}; its implications are discussed in Sect.~\ref{sec:limitations}.

Section~\ref{sec:data} defines the data and forward models. Section~\ref{sec:results} presents the likelihood and posterior comparisons, recovery experiments, and sensitivities. Section~\ref{sec:conclusions} summarises their interpretation. The appendices document analytic normalisation, covariance propagation, numerical verification, and reproducibility.

\section{Data and forward modelling}\label{sec:data}
\subsection{Lens sample and survey provenance}
\label{sec:lenses}

Our parent catalogue consists of 130 lenses from the compilation of \citet{stz1902} for which individual photometric slopes were provided by Y.~Chen. We use the subset with the required photometric and kinematic measurements. The observations originate from the Sloan Lens Advanced Camera for Surveys (SLACS) Survey, SLACS for the Masses (S4TM), the Baryon Oscillation Spectroscopic Survey (BOSS) Emission-Line Lens Survey (BELLS), and its Galaxy--Ly$\alpha$ Emitter Systems extension (BELLS GALLERY). Table~\ref{tab:surveys} lists the parent and matched counts with the original survey references. SLACS and S4TM dominate the matched sample, which retains only one BELLS GALLERY lens.

\begin{table*}[t]
\centering
\caption{Provenance of the parent catalogue and the baseline matched sample. Counts refer to this analysis, not to the total size of each original survey.}
\label{tab:surveys}
\begin{tabular}{lrrl}
\toprule
Survey & Parent & Matched & Original observational references \\
\midrule
SLACS & 57 & 53 & \citet{Bolton2006,Bolton2008,Auger2009,Auger2010} \\
S4TM & 38 & 36 & \citet{Shu2015,Shu2017} \\
BELLS & 21 & 17 & \citet{Brownstein2012} \\
BELLS GALLERY & 14 & 1 & \citet{Shu2016a,Shu2016b} \\
\midrule
Total & 130 & 107 & \\
\bottomrule
\end{tabular}
\end{table*}

The lens observables are the lens and source redshifts $(z_l,z_s)$, Einstein radius $\theta_E$, aperture radius $\theta_{\rm ap}$, effective light radius $\theta_{\rm eff}$, aperture velocity dispersion $\sigma_{\rm ap}$ and its quoted statistical uncertainty, and the individual luminosity slope $\delta_i$. We use the catalogue measurements directly. The photometric measurements use Hubble Space Telescope (HST) F814W imaging for SLACS, S4TM, and BELLS, and F606W imaging for BELLS GALLERY, as described by \citet{stz1902}.

We use the individual aperture and effective radii listed in the catalogue. We interpret $\theta_{\rm ap}$ as a circular aperture radius and predict the second moment within that aperture. The listed dispersions agree with Table A1 of \citet{stz1902}, before that paper's transformation to $R_e/2$. Direct modelling of the original aperture therefore avoids applying the transformation twice. The heterogeneous spectroscopic extraction and light weighting remain relevant to the sensitivity tests.

\subsection{Supernova distance ratios and matching}\label{sec:pairing}

Matched-distance tests require observables at comparable redshifts. Fixed redshift windows have been used in cluster--supernova and lens--supernova distance-duality tests \citep{Holanda2010,Liao2016}. A fixed redshift difference corresponds to different fractional distance differences across the sample. \citet{Zhou2021} introduced a distance-deviation consistency criterion, which was subsequently applied to lens-density slopes \citep{Hu2023} and supernova-magnitude calibration \citep{Hu2023p}. These applications motivate a relative distance tolerance rather than a universal redshift window.

The distance metric and the allocation rule are separate choices. \citet{Hu2026cluster} used a relative comoving-distance tolerance for cluster--supernova pairs and examined 3\% and 5\% thresholds, with a greedy allocation without replacement. A related lens--quasar analysis used a deterministic maximum-cardinality, minimum-mismatch assignment \citep{Hu2026curvature}. Here we impose global uniqueness across the two supernova roles and use the two-stage optimisation described below. An alternative strategy reconstructs a continuous distance function, for example with Gaussian processes \citep{Seikel2012}; our matched-catalogue construction instead retains the selected supernova measurements and their covariance blocks.
We use the corrected apparent magnitudes $m_b$ and full statistical-plus-systematic covariance from the official \pantheon\ release \citep{Brout2022,Scolnic2022}. We use the supplied \textsc{CosmoMatcher} version 1.4.3 implementation\footnote{\url{https://github.com/HUJIAN0000/CosmoMatcher}} to match supernovae to the lens and source redshifts using $z_{\rm HD}$ and a fractional comoving-distance tolerance of 0.05 in a flat fiducial background with $\Omega_m=0.30$. The two-stage mixed-integer linear programme first maximises the number of complete systems and then minimises squared fractional mismatch. Each identifier in the release's CID field is used once globally. A catalogue-order tie break selects a reproducible assignment. The primary sample contains 107 lenses and 214 distinct CIDs. We extract all covariance blocks using the original supernova row indices. The matching metric and fiducial background specify the assignment, while the dynamical fit uses the resulting supernova distance ratios.

For supernovae at the target lens and source redshifts, spatial flatness and distance duality give \citep{Rasanen2015,Liao2016}
\begin{equation}\label{eq:Robs}
R\equiv\frac{D_{ls}}{D_s}=1-\frac{\chi_l}{\chi_s}
=1-10^{0.2(m_{b,l}-m_{b,s})}\frac{1+z_s}{1+z_l}.
\end{equation}
The common standardised absolute magnitude and an overall distance scale cancel. Differential calibration and redshift-dependent standardisation remain relevant. In practice, the matched supernova redshifts differ from the target redshifts, while the explicit factors in Eq.~(\ref{eq:Robs}) use the target values. The resulting $R_{\rm SN}$ is therefore a finite-matching approximation. We assess finite matching through alternative tolerances, an input-order control, and model-assisted redshift corrections (Sect.~\ref{sec:distancechecks}).

\subsection{Common mass model and anisotropy}\label{sec:P2}
The baseline two-coefficient (P2) mass model is spherical with $\rho(r)\propto r^{-\gamma}$ and
\begin{equation}\label{eq:P2}
\gamma_i=\gamma_0+\gamma_z z_{l,i}.
\end{equation}
The intercept $\gamma_0$ gives the extrapolated zero-redshift value of the sample relation, distinct from a radial gradient within a lens. We also report $\gamma_p=\gamma_0+z_p\gamma_z$ at the sample median $z_p=0.1959$. This is a transformation of each joint posterior sample, without changing the original prior.

The constant anisotropy is $\beta\equiv\beta_{\rm ani}=1-(\sigma_\theta^2+\sigma_\phi^2)/(2\sigma_r^2)$, shared by the sample. The hypotheses are a free common $\beta$, fixed isotropy, and fixed $\beta=0.18$. Shared spatial constancy is a restrictive model, distinct from a population distribution with intrinsic anisotropy scatter.

\subsection{Power-law and curved tracers}\label{sec:tracers}
For the power-law tracer, $\nu_i(r)\propto r^{-\delta_i}$. The individual slopes were obtained by fitting a PSF-convolved projected profile $I(R)\propto R^{1-\delta_i}$ within $\theta_{\rm eff}/2$ \citep{stz1902}. In 92 of our 107 lenses, the spectroscopic aperture extends beyond this photometric fitting radius: the median aperture-to-fit-radius ratio is 1.53 and the maximum is 4.55. A global power-law tracer thus extrapolates the fitted inner profile over the region contributing to the aperture dispersion. \citet{stz1902} report relative statistical slope errors below 1\%, while their population dispersion 0.085 describes variation between galaxies. Individual error estimates and imaging-fit records are unavailable for the supplied slopes. The analytic aperture prediction is
\begin{equation}\label{eq:sigma_th}
\sigma_{\rm th}^2=\frac{c^2}{2\sqrt\pi}\frac{\theta_E}{R_{\rm SN}}
 F(\gamma,\delta_i,\beta)\left(\frac{\theta_{\rm ap}}{\theta_E}\right)^{2-\gamma},
\end{equation}
where $c$ is the speed of light, angular radii in the prefactor are in radians, and $F$ is given in Appendix~\ref{sec:analytic}.

For comparison, we use a Hernquist tracer \citep{Hernquist1990}, with projected half-light radius $R_e$ corresponding to the catalogue angular radius $\theta_{\rm eff}$,
\begin{equation}
\nu(r)\propto\frac{1}{r(r+a)^3},\qquad a=R_e/1.81527,
\end{equation}
and the exact Abel deprojection of the projected S\'ersic form
\begin{equation}
I(R)\propto\exp[-b_n(R/R_e)^{1/n}],\qquad
\nu(r)=-\frac1\pi\int_r^\infty\frac{\mathrm d I/\mathrm d R}{\sqrt{R^2-r^2}}\,\mathrm dR.
\end{equation}
The constant $b_n$ imposes the projected half-light condition \citep{GrahamDriver2005}. We use $n=4$ for the main S\'ersic comparison and $n=2$ and 6 as sensitivity scenarios. These models use the measured effective radii and assumed profile families, replacing the individual inner-slope constraint used by the power-law model. The comparison therefore tests the overall photometric representation. A common imaging likelihood would be required to isolate curvature while retaining identical photometric information.

For each curved tracer, we solve the spherical Jeans equation \citep{Binney1980,MamonLokas2005},
\begin{equation}
\nu(r)\sigma_r^2(r)=r^{-2\beta}\int_r^\infty\nu(s)\frac{GM(s)}{s^2}s^{2\beta}\,\mathrm ds,
\end{equation}
normalising the power-law mass by its Einstein projected mass and the stated $R_{\rm SN}$. We then project the pressure and integrate the line-of-sight second moment over the circular aperture. Both the moment numerator and light denominator use the same aperture weighting. The baseline has no PSF convolution; additional cases use an assumed circular Gaussian PSF with $\sigma_{\rm PSF}=0.64$ arcsec. This assumed seeing value defines a PSF sensitivity scenario for the catalogue.

We evaluate the numerical integrals on 4096 logarithmic radial nodes, with 128 Abel and 192 angular quadrature nodes. A Chebyshev representation accelerates repeated evaluations. Appendix~\ref{sec:numerical} describes direct-quadrature comparisons, independent off-grid tests, and checks at doubled resolution.

\subsection{Conditional likelihood and parameter domain}\label{sec:stats}
In matched-catalogue order, the magnitude-difference covariance is
\begin{equation}\label{eq:Cdifference}
\mathbf C_{\Delta m}=\mathbf C_{ll}+\mathbf C_{ss}-\mathbf C_{ls}-\mathbf C_{ls}^{T}.
\end{equation}
First-order derivatives propagate it to $\mathbf C_R$ and then to $\mathbf C_{\sigma,{\rm SN}}$ (Appendix~\ref{sec:propagation}). For the dispersion residual $\mathbf D=\boldsymbol\sigma_{\rm ap}-\boldsymbol\sigma_{\rm th}$,
\begin{equation}\label{eq:Ctotal}
\mathbf C_{\rm tot}=\mathbf C_{\sigma,{\rm SN}}+
\operatorname{diag}\!\left[\sigma_{\rm stat}^2+(0.03\sigma_{\rm ap})^2+(f_{\rm int}\sigma_{\rm ap})^2\right],
\end{equation}
and
\begin{equation}\label{eq:likelihood}
\ln\mathcal L=-\tfrac12[\mathbf D^T\mathbf C_{\rm tot}^{-1}\mathbf D+
\ln\det\mathbf C_{\rm tot}+N\ln(2\pi)].
\end{equation}
The propagated covariance is recomputed as the model changes, with all cross-system terms retained and Cholesky factorisation used for evaluation. The 3\% term is phenomenological and does not explicitly model coherent calibration or geometry. The fractional-scatter parameter $f_{\rm int}$ is distinct from $\delta_i$. Its observational dispersion anchor is held fixed during parameter inference.

This is a conditional Gaussian approximation, not a latent-supernova likelihood. In particular, the nonlinear transformation of uncertain magnitudes can be non-Gaussian when $R$ is poorly constrained. Section~\ref{sec:distancechecks} quantifies the sensitivity to nonlinear propagation in poorly constrained systems.

The prior boxes are $1<\gamma_0<3$, $-2<\gamma_z<2$, $-1<\beta<1$, and $0<f_{\rm int}<1$, with $1.05<\gamma_i<2.95$ at every lens. The Gaussian free-$\beta$ comparison replaces the uniform beta density by $\mathcal N(0.18,0.13^2)$ truncated to $(-1,1)$. The historical implementation in \citet{stz1902} instead restricted this Gaussian to its mean plus or minus two standard deviations, $[-0.08,0.44]$. Our broad-support Gaussian comparison is consequently not an exact reproduction of that historical prior. For the analytic global power-law moments we additionally impose $1<\delta_i<3$, $1<\xi_i<3$, and $\xi_i>2\beta$, where $\xi_i=\gamma_i+\delta_i-2$. These are convergence conditions, not a proof of a non-negative phase-space distribution function. The curved-tracer integrals are assessed with their own finite-moment domain, without imposing the global power-law $\xi$ restrictions.

We determine each maximum-likelihood estimate (MLE) with three independent differential-evolution searches followed by Nelder--Mead refinement. The maxima must agree within 0.05 in $\ln\mathcal L$ and satisfy the nested likelihood ordering. Each main posterior is sampled with four independent \texttt{emcee} ensembles \citep{ForemanMackey2013} of 32 walkers, discarding 3000 initial steps. We extend the ensembles according to the convergence criteria in Appendix~\ref{sec:numerical}; accepted lengths are listed in Table~\ref{tab:diagnostics}.

For $k$ fitted parameters, the Akaike information criterion (AIC; \citealt{Akaike1974}) is $\mathrm{AIC}=2k-2\ln\mathcal L_{\max}$ and the Bayesian information criterion (BIC; \citealt{Schwarz1978}) is $\mathrm{BIC}=k\ln N-2\ln\mathcal L_{\max}$. We report
\begin{equation}
\Delta\mathrm{BIC}_{b}=\mathrm{BIC}_{\beta=b}-\mathrm{BIC}_{\rm free},
\end{equation}
so positive values favour the free model under this criterion. BIC with $N$ lens systems is an approximate description for these correlated, model-covariance-dependent data, not a calibrated significance or probability. Flat and Gaussian free-$\beta$ priors share the likelihood maximum and hence the likelihood-based criteria. We compare criteria only within the same data set.

\section{Results and discussion}\label{sec:results}
\subsection{Tracer-dependent likelihood comparisons}\label{sec:comparison}
\begin{figure*}[t]\centering\includegraphics[width=.9\textwidth]{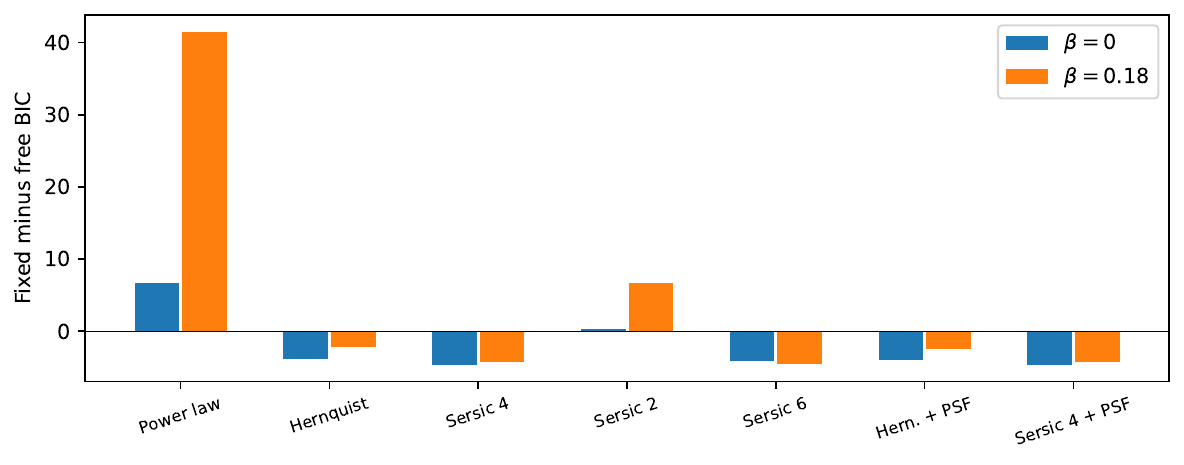}
\caption{Tracer and PSF dependence of the fixed-minus-free BIC differences.}\label{fig:bic}\end{figure*}

Table~\ref{tab:tracer} and Fig.~\ref{fig:bic} compare likelihood maxima on the newly matched 107-lens catalogue. The individual-slope power-law tracer yields $\Delta\mathrm{BIC}_0=6.74$ and $\Delta\mathrm{BIC}_{0.18}=41.51$. The corresponding values are $-3.89$ and $-2.26$ for Hernquist, and $-4.67$ and $-4.30$ for S\'ersic $n=4$. The criterion preference for freeing the common anisotropy therefore depends on the assumed light profile. 

The isotropic S\'ersic $n=4$ likelihood is $-513.79$, compared with $-514.03$ for the free power-law case, despite fitting one fewer parameter. The $n=2$ case retains $\Delta\mathrm{BIC}_{0.18}=6.61$, while its isotropy comparison is only 0.24. For $n=6$ the differences are $-4.16$ and $-4.55$. The free MLE beta changes from $-0.54$ at $n=2$ to $-0.01$ at $n=4$ and $+0.31$ at $n=6$. A difference near $-\ln107=-4.67$ corresponds to nearly equal likelihood maxima, with the parameter-count penalty favouring the fixed case.
\begin{table*}[t]\centering\small
\caption{P2 likelihood maxima. Differences are fixed minus free within each tracer scenario on the same 107 lenses. Pivot slopes use $z_p=0.1959$. PSF and S\'ersic-index choices are assumed scenarios.}\label{tab:tracer}\begin{tabular}{llrrrrr}
\toprule
Tracer & $\beta$ treatment & $\ln\mathcal L_{\max}$ & $\Delta\mathrm{BIC}$ & $\widehat\gamma_p$ & $\widehat\beta$ & $\widehat f_{\rm int}$ \\
\midrule
Power law & Free, flat & -514.03 & 0.00 & 2.018 & -0.569 & 0.080 \\
Power law & Fixed 0 & -519.74 & 6.74 & 1.810 & 0.000 & 0.086 \\
Power law & Fixed 0.18 & -537.12 & 41.51 & 1.740 & 0.180 & 0.107 \\
Hernquist & Free, flat & -514.37 & 0.00 & 2.038 & -0.331 & 0.081 \\
Hernquist & Fixed 0 & -514.77 & -3.89 & 1.987 & 0.000 & 0.082 \\
Hernquist & Fixed 0.18 & -515.58 & -2.26 & 1.948 & 0.180 & 0.083 \\
S\'ersic $n=4$ & Free, flat & -513.79 & 0.00 & 1.995 & -0.012 & 0.080 \\
S\'ersic $n=4$ & Fixed 0 & -513.79 & -4.67 & 1.993 & 0.000 & 0.080 \\
S\'ersic $n=4$ & Fixed 0.18 & -513.98 & -4.30 & 1.957 & 0.180 & 0.081 \\
S\'ersic $n=2$ & Free, flat & -514.00 & 0.00 & 2.019 & -0.541 & 0.080 \\
S\'ersic $n=2$ & Fixed 0 & -516.45 & 0.24 & 1.881 & 0.000 & 0.084 \\
S\'ersic $n=2$ & Fixed 0.18 & -519.64 & 6.61 & 1.808 & 0.180 & 0.089 \\
S\'ersic $n=6$ & Free, flat & -513.61 & 0.00 & 1.977 & 0.309 & 0.080 \\
S\'ersic $n=6$ & Fixed 0 & -513.86 & -4.16 & 2.021 & 0.000 & 0.080 \\
S\'ersic $n=6$ & Fixed 0.18 & -513.67 & -4.55 & 1.999 & 0.180 & 0.080 \\
Hernquist + PSF & Free, flat & -513.99 & 0.00 & 2.029 & -0.282 & 0.080 \\
Hernquist + PSF & Fixed 0 & -514.30 & -4.05 & 1.984 & 0.000 & 0.081 \\
Hernquist + PSF & Fixed 0.18 & -515.07 & -2.51 & 1.943 & 0.180 & 0.082 \\
S\'ersic $n=4$ + PSF & Free, flat & -513.57 & 0.00 & 1.991 & -0.003 & 0.080 \\
S\'ersic $n=4$ + PSF & Fixed 0 & -513.57 & -4.67 & 1.991 & 0.000 & 0.080 \\
S\'ersic $n=4$ + PSF & Fixed 0.18 & -513.75 & -4.33 & 1.955 & 0.180 & 0.080 \\
\bottomrule
\end{tabular}
\end{table*}
\subsection{Slope shifts and posterior dependence}\label{sec:posteriors}
The power-law MLE zero-redshift intercept changes from 2.194 with free beta to 2.019 and 1.962 at fixed 0 and 0.18. For Hernquist the values are 2.186, 2.125, and 2.076; for S\'ersic $n=4$ they are 2.128, 2.126, and 2.084. At the pivot redshift the power-law values are 2.018, 1.810, and 1.740, compared with 1.995, 1.993, and 1.957 for S\'ersic $n=4$. Thus the sensitivity is also present at a redshift represented in the sample.

\begin{figure*}[t]\centering\includegraphics[width=.9\textwidth]{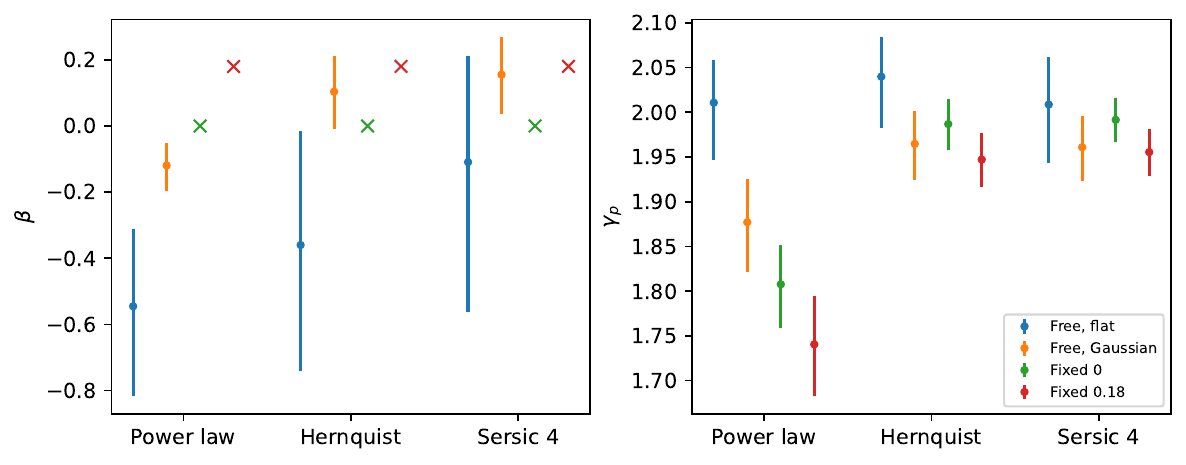}\caption{Effective beta and pivot-slope posteriors versus tracer and anisotropy treatment. Crosses denote fixed beta values.}\label{fig:posterior}\end{figure*}

\begin{figure*}[t]\centering\includegraphics[width=.78\textwidth]{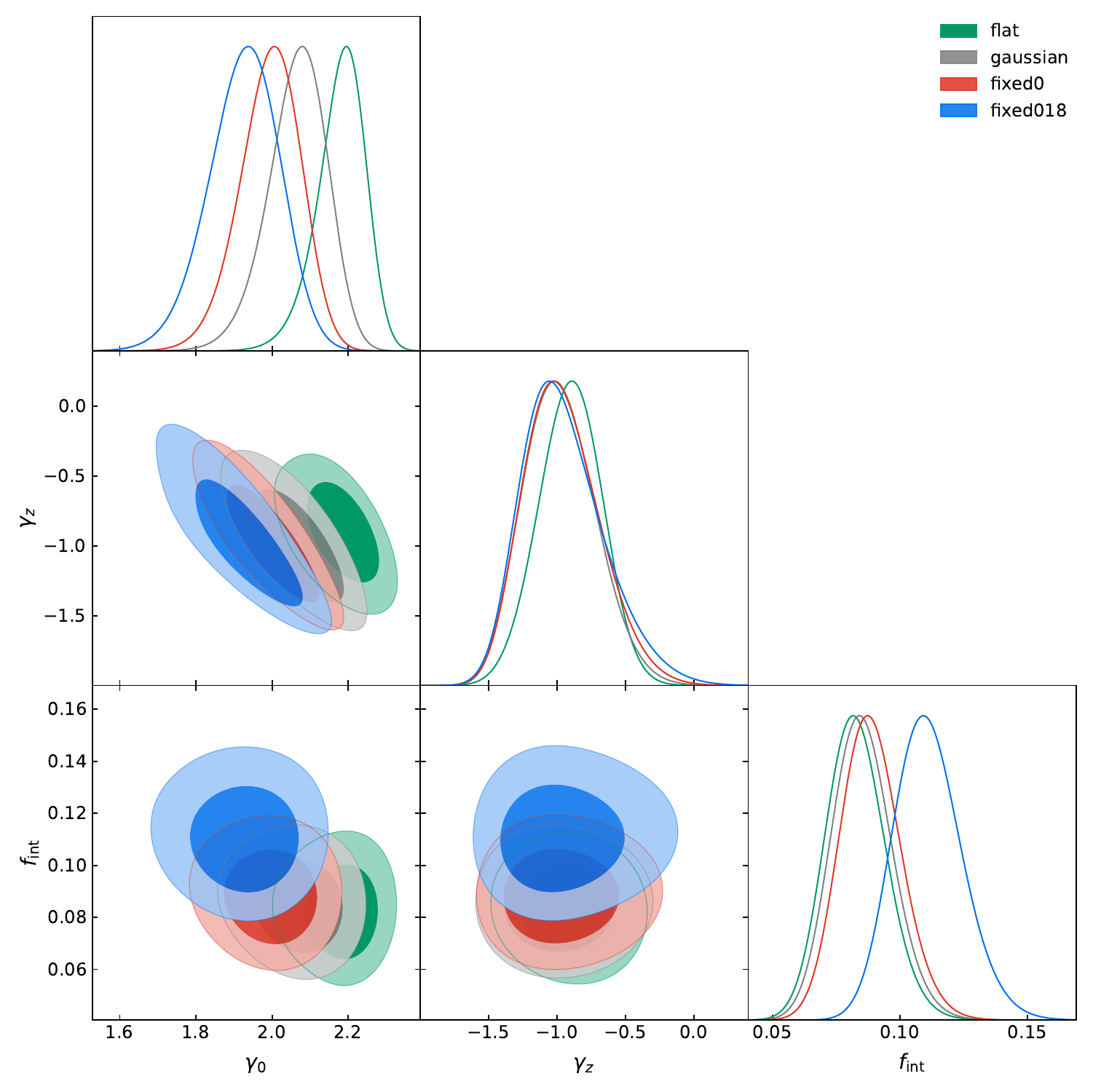}\caption{Power-law P2 posterior comparison under all four anisotropy treatments. Only common sampled parameters are displayed; free beta is marginalised over.}\label{fig:p2_joint_comparison}\end{figure*}

Table~\ref{tab:posterior} and Fig.~\ref{fig:posterior} give posterior medians and central 68\% intervals under both free-beta priors and both fixed hypotheses. Every pivot interval is computed from the joint draws. Figure~\ref{fig:p2_joint_comparison} shows the common parameters across the four power-law treatments. Appendix~\ref{sec:joint_posteriors} shows the complete joint posterior distributions. The free-flat power-law posterior has 1.48\% of beta draws within 0.02 of its lower prior bound $-1$.
\begin{table*}[t]\centering\small\caption{P2 posterior medians and central 68\% intervals. Fixed beta entries have no posterior interval.}\label{tab:posterior}\begin{tabular}{llrrrrr}
\toprule
Tracer & $\beta$ treatment & $\gamma_p$ & $\gamma_z$ & $\beta$ & $f_{\rm int}$ & $\gamma_0$ \\
\midrule
Power law & Free, flat & $2.011_{-0.064}^{+0.047}$ & $-0.901_{-0.234}^{+0.225}$ & $-0.545_{-0.271}^{+0.234}$ & $0.082_{-0.011}^{+0.012}$ & $2.186_{-0.068}^{+0.055}$ \\
Power law & Free, Gaussian & $1.877_{-0.055}^{+0.049}$ & $-0.996_{-0.251}^{+0.276}$ & $-0.120_{-0.076}^{+0.067}$ & $0.085_{-0.011}^{+0.012}$ & $2.071_{-0.081}^{+0.071}$ \\
Power law & Fixed 0 & $1.808_{-0.048}^{+0.044}$ & $-0.987_{-0.253}^{+0.298}$ & $0$ & $0.088_{-0.011}^{+0.012}$ & $1.999_{-0.083}^{+0.074}$ \\
Power law & Fixed 0.18 & $1.740_{-0.057}^{+0.054}$ & $-0.994_{-0.257}^{+0.337}$ & $0.18$ & $0.110_{-0.013}^{+0.014}$ & $1.930_{-0.096}^{+0.088}$ \\
Hernquist & Free, flat & $2.040_{-0.057}^{+0.044}$ & $-0.749_{-0.193}^{+0.196}$ & $-0.360_{-0.381}^{+0.345}$ & $0.084_{-0.011}^{+0.012}$ & $2.186_{-0.077}^{+0.065}$ \\
Hernquist & Free, Gaussian & $1.965_{-0.040}^{+0.037}$ & $-0.670_{-0.202}^{+0.204}$ & $0.104_{-0.113}^{+0.109}$ & $0.085_{-0.011}^{+0.012}$ & $2.096_{-0.064}^{+0.061}$ \\
Hernquist & Fixed 0 & $1.987_{-0.029}^{+0.028}$ & $-0.698_{-0.196}^{+0.198}$ & $0$ & $0.084_{-0.011}^{+0.012}$ & $2.123_{-0.053}^{+0.052}$ \\
Hernquist & Fixed 0.18 & $1.947_{-0.030}^{+0.030}$ & $-0.652_{-0.203}^{+0.205}$ & $0.18$ & $0.085_{-0.011}^{+0.012}$ & $2.075_{-0.056}^{+0.055}$ \\
S\'ersic $n=4$ & Free, flat & $2.009_{-0.066}^{+0.053}$ & $-0.682_{-0.184}^{+0.188}$ & $-0.109_{-0.454}^{+0.322}$ & $0.083_{-0.011}^{+0.012}$ & $2.143_{-0.083}^{+0.068}$ \\
S\'ersic $n=4$ & Free, Gaussian & $1.961_{-0.038}^{+0.035}$ & $-0.652_{-0.190}^{+0.193}$ & $0.155_{-0.118}^{+0.113}$ & $0.083_{-0.011}^{+0.012}$ & $2.089_{-0.059}^{+0.055}$ \\
S\'ersic $n=4$ & Fixed 0 & $1.992_{-0.025}^{+0.024}$ & $-0.676_{-0.186}^{+0.184}$ & $0$ & $0.082_{-0.011}^{+0.012}$ & $2.124_{-0.047}^{+0.046}$ \\
S\'ersic $n=4$ & Fixed 0.18 & $1.955_{-0.026}^{+0.026}$ & $-0.648_{-0.193}^{+0.192}$ & $0.18$ & $0.083_{-0.011}^{+0.012}$ & $2.082_{-0.050}^{+0.049}$ \\
\bottomrule
\end{tabular}
\end{table*}
\subsection{Controlled recovery and tracer mismatch}\label{sec:mocks}
We generate correlated dispersion vectors for four MLE-based generating models: isotropic power law, free-beta power law, isotropic S\'ersic $n=4$, and isotropic Hernquist. There are 200 predetermined realisations per truth. Each vector is fitted with free beta, fixed 0, and fixed 0.18, under both its true tracer and an alternative family. All 800 realisations and their 4800 fitted models pass the specified optimisation checks. Null and alternative experiments use the same noise prescription. The generating dispersion sets the fractional-scatter variance anchor, which remains fixed during fitting. The conditional noise model is held fixed across these paired comparisons.

\begin{figure*}[t]\centering\includegraphics[width=.9\textwidth]{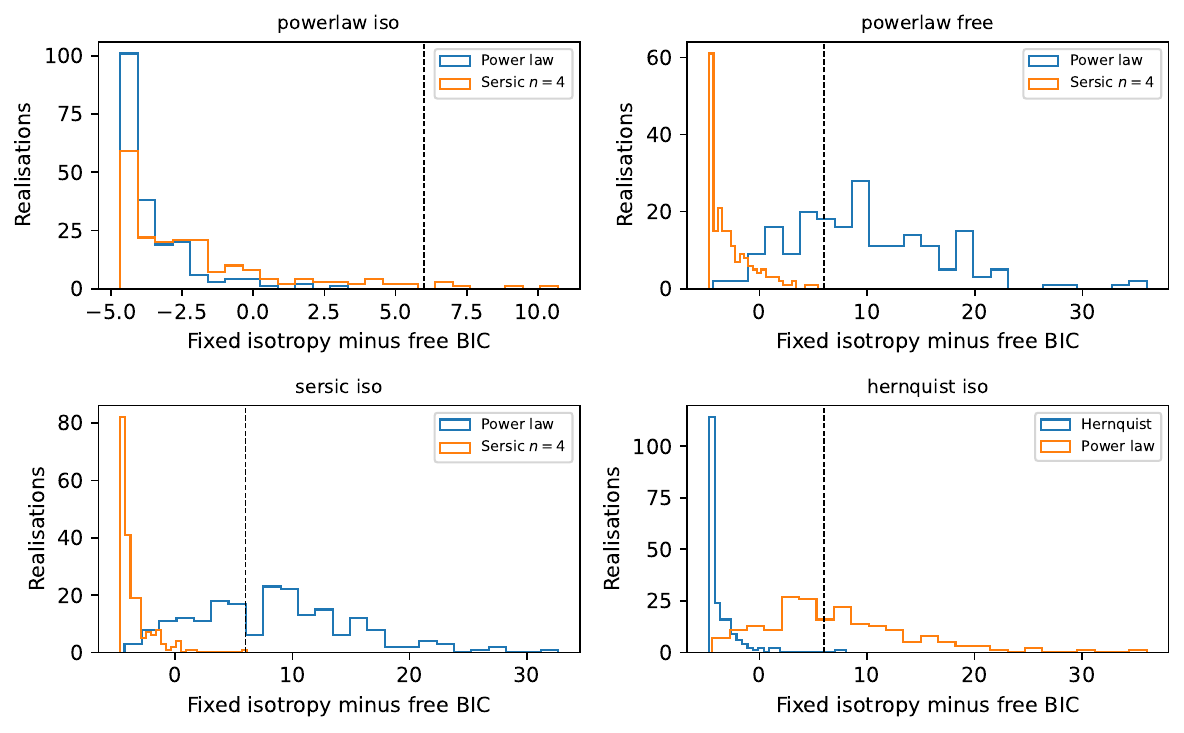}\caption{Truth-matched and alternative-tracer recovery distributions for the same simulated vectors. The line at 6 is a descriptive BIC threshold, not a universal significance.}\label{fig:mocks}\end{figure*}

Table~\ref{tab:mocks} and Fig.~\ref{fig:mocks} report paired comparisons. Under isotropic S\'ersic $n=4$ and Hernquist truths, fitting the power-law tracer produces $\Delta\mathrm{BIC}_0>6$ in 60.0\% and 50.0\% of realisations, respectively, with mean effective-beta shifts of $-0.547$ and $-0.492$. Matched-tracer isotropic fits cross this threshold in only 0.5\% and 1.0\% of those realisations. Under an isotropic power-law truth, the matched power-law rate is 0.5\%. The free-beta power-law truth gives 67.5\% recovery above the threshold under its own family. These frequencies depend on the chosen MLE truths, noise model, catalogue, and threshold. The cross-tracer fits quantify how the specified mismatch can generate an effective anisotropy preference.
\begin{table*}[t]\centering\small\caption{Conditional recovery experiments, 200 realisations per truth. Threshold frequencies include Wilson 95\% intervals in percent. Biases refer to each stated generative truth.}\label{tab:mocks}\begin{tabular}{llrrrr}
\toprule
Generating truth & Fitted tracer & $\langle\Delta\mathrm{BIC}_0\rangle$ & $P(\Delta\mathrm{BIC}_0>6)$ (\%) & $\langle\widehat\beta-\beta_{\rm true}\rangle$ & $\langle\widehat\gamma_p-\gamma_{p,{\rm true}}\rangle$ \\
\midrule
Hernquist, isotropic & Hernquist & -3.63 & 1.0 [0.3, 3.6] & -0.047 & -0.008 \\
Hernquist, isotropic & Power law & 6.88 & 50.0 [43.1, 56.9] & -0.492 & -0.008 \\
Power law, free MLE & Power law & 9.81 & 67.5 [60.7, 73.6] & -0.029 & -0.007 \\
Power law, free MLE & S\'ersic $n=4$ & -2.69 & 0.0 [0.0, 1.9] & 0.026 & 0.036 \\
Power law, isotropic & Power law & -3.56 & 0.5 [0.1, 2.8] & -0.005 & -0.010 \\
Power law, isotropic & S\'ersic $n=4$ & -1.99 & 3.0 [1.4, 6.4] & 0.321 & 0.125 \\
S\'ersic $n=4$, isotropic & Power law & 8.44 & 60.0 [53.1, 66.5] & -0.547 & 0.000 \\
S\'ersic $n=4$, isotropic & S\'ersic $n=4$ & -3.56 & 0.5 [0.1, 2.8] & -0.085 & -0.006 \\
\bottomrule
\end{tabular}
\end{table*}
\subsection{Distance uncertainty and matching}\label{sec:distancechecks}
The linearised fractional distance-ratio uncertainty has median 0.0651, reaches 0.443, and exceeds 0.20 in 12 systems. In 50000 correlated Gaussian magnitude-difference draws, the largest fraction yielding $R\le0.001$ is 3.6\% (SDSSJ2122+0409). These diagnostic draws quantify nonlinear propagation about the observed magnitudes; their interpretation is discussed in Sect.~\ref{sec:limitations}.

Excluding systems above 20\% linearised fractional distance-ratio uncertainty leaves 95 lenses. The power-law isotropy comparison becomes 2.51; Hernquist and S\'ersic $n=4$ give $-4.04$ and $-4.54$. Exclusion changes both uncertainty and sample design, so the difference cannot be attributed uniquely to nonlinear propagation. The likelihood approximation and its implications are discussed in Sect.~\ref{sec:limitations}.

Appendix~\ref{sec:sensitivity} gives corrections for finite redshift mismatch under smooth flat backgrounds with $\Omega_m=0.25$, 0.30, and 0.35, together with a heliocentric-redshift transfer check. These corrections transform both the ratios and covariance. The dependence on the assumed background is included in these comparisons.

We also repeat the two-stage matching at tolerances 0.03 and 0.07, obtaining 104 and 111 lenses. Their power-law isotropy comparisons are 3.69 and 6.90. Reversing the SN input ordering gives another 107-lens catalogue with 78 complete triplets shared with the primary assignment and differences of 6.51 and 41.67 for power law. All three alternatives retain negative isotropy differences for Hernquist and S\'ersic $n=4$ (Table~\ref{tab:matching}). The principal tracer dependence persists across these assignments, while the strength of the power-law preference varies.
\subsection{Photometric uncertainty and structural control}\label{sec:otherchecks}
Coherent offsets of each measured $\delta_i$ by $\pm0.05$ and $\pm0.10$ are refitted on the same primary catalogue (Table~\ref{tab:sensitivity}). These assumed offsets probe sensitivity to photometric calibration. We separately integrate over latent slopes with a truncated-normal prior on $(1.001,2.999)$ and enforce the dynamical convergence domain jointly with the global parameters. Four independent chains use blocked elliptical-slice updates for Gaussian latent coordinates mapped through the truncated-normal cumulative distribution, and global random-walk updates adapted only during burn-in \citep{Murray2010}.

The smallest-error scenario uses $\sigma_{\delta,i}=0.01\widehat\delta_i$, motivated by the upper bound on relative statistical errors stated by \citet{stz1902}. We treat this width as a statistical upper-bound scenario. Additional absolute widths 0.05 and 0.10, and width 0.05 with common offset $b\sim\mathcal N(0,0.05^2)$, probe broader modelling uncertainties. The published population scatter 0.085 is not substituted for a measurement error. Table~\ref{tab:deltapost} gives the new results. All eight posterior calculations satisfy the diagnostics for the global parameters and every varying latent coordinate. These slope-error priors apply to the power-law tracer; we report the marginal posteriors separately from the likelihood-based BIC comparisons.
\begin{table*}[t]\centering\small\caption{Power-law posterior sensitivity to slope-error scenarios. Entries are medians and central 68\% intervals. The relative-1\% row is a statistical upper-bound scenario; the other widths are assumed sensitivity inputs.}\label{tab:deltapost}\begin{tabular}{llrrrr}
\toprule
Slope-error scenario & $\beta$ treatment & $\gamma_p$ & $\gamma_z$ & $\beta$ & $f_{\rm int}$ \\
\midrule
$\sigma_\delta=0.05$ & Free, flat & $2.008_{-0.065}^{+0.048}$ & $-0.904_{-0.235}^{+0.225}$ & $-0.533_{-0.275}^{+0.231}$ & $0.082_{-0.011}^{+0.012}$ \\
$\sigma_\delta=0.05$ & Fixed 0 & $1.810_{-0.047}^{+0.043}$ & $-1.000_{-0.254}^{+0.296}$ & $0$ & $0.085_{-0.011}^{+0.012}$ \\
$\sigma_\delta=0.10$ & Free, flat & $2.001_{-0.064}^{+0.050}$ & $-0.909_{-0.241}^{+0.226}$ & $-0.499_{-0.278}^{+0.219}$ & $0.080_{-0.011}^{+0.012}$ \\
$\sigma_\delta=0.10$ & Fixed 0 & $1.816_{-0.045}^{+0.041}$ & $-1.022_{-0.255}^{+0.294}$ & $0$ & $0.076_{-0.013}^{+0.013}$ \\
$\sigma_\delta=0.05$, $\sigma_b=0.05$ & Free, flat & $2.007_{-0.065}^{+0.048}$ & $-0.914_{-0.256}^{+0.237}$ & $-0.540_{-0.267}^{+0.226}$ & $0.082_{-0.011}^{+0.012}$ \\
$\sigma_\delta=0.05$, $\sigma_b=0.05$ & Fixed 0 & $1.866_{-0.051}^{+0.044}$ & $-0.954_{-0.251}^{+0.270}$ & $0$ & $0.084_{-0.011}^{+0.012}$ \\
$\sigma_{\delta,i}=0.01\widehat\delta_i$ & Free, flat & $2.010_{-0.064}^{+0.048}$ & $-0.904_{-0.232}^{+0.225}$ & $-0.541_{-0.272}^{+0.232}$ & $0.082_{-0.011}^{+0.012}$ \\
$\sigma_{\delta,i}=0.01\widehat\delta_i$ & Fixed 0 & $1.808_{-0.048}^{+0.044}$ & $-0.990_{-0.252}^{+0.297}$ & $0$ & $0.088_{-0.011}^{+0.012}$ \\
\bottomrule
\end{tabular}
\end{table*}
The effective-beta medians and central 68\% intervals under the free-flat prior are $\sigma_\delta=0.05$: $-0.533_{-0.275}^{+0.231}$; $\sigma_\delta=0.10$: $-0.499_{-0.278}^{+0.219}$; $\sigma_\delta=0.05$, $\sigma_b=0.05$: $-0.540_{-0.267}^{+0.226}$; $\sigma_{\delta,i}=0.01\widehat\delta_i$: $-0.541_{-0.272}^{+0.232}$. The four rows quantify sensitivity to the stated slope-error models.

Removing the BELLS family leaves 89 lenses and gives $\Delta\mathrm{BIC}_0=7.40$ for power law; removing S4TM leaves 71 lenses and gives $-1.82$. The main curved-tracer comparisons remain negative. Survey, redshift, and aperture coverage change together, preventing causal attribution to a single survey.

As a three-coefficient structural control, P3 adds $\gamma_s\log_{10}\widetilde\Sigma_L$ to Eq.~(\ref{eq:P2}), where
\begin{equation}
\widetilde\Sigma_L=166\,\frac{D_s/(1\,\mathrm{Mpc})}{[D_l/(1\,\mathrm{Mpc})][D_{ls}/(1\,\mathrm{Mpc})]}
\left(\frac{\theta_E}{\theta_{\rm eff}}\right)^2.
\end{equation}
The predictor distances assume flat $\Omega_m=0.30$ and $H_0=70\,\kms\,\mathrm{Mpc}^{-1}$. This differs from the kinematic predictor used by \citet{stz1902}. The power-law P3 comparisons give 6.87 and 35.19. Its pivot slope additionally conditions on the sample median predictor, and its posterior parameters appear in Table~\ref{tab:p3posterior}.
\subsection{Scope and limitations}\label{sec:limitations}
All models assume spherical symmetry, a power-law total mass density, and one radially constant beta shared by the lenses. They do not fit flattening, rotation, anisotropy gradients, or population scatter. Integral-field surveys resolve a variety of kinematic structures and photometric--kinematic alignments in early-type galaxies \citep{Krajnovic2011}. The curved tracers replace the individual inner-slope information with effective radii and assumed profile families. Original imaging and consistent PSF modelling are needed to compare profiles under a common photometric likelihood. A broad-band light profile can also differ from the spectral weighting of the kinematic measurement \citep{Fores2026}.

The SN ratios are correlated and use finite redshift matching and first-order covariance propagation; Sect.~\ref{sec:distancechecks} reports the corresponding sensitivity tests. Gaussian magnitude-difference draws describe nonlinear propagation around the observed magnitudes, while a latent-distance analysis would require a specified prior and joint treatment of the supernova data. The additional 3\% variance term and fitted scatter give a phenomenological uncertainty model. The fitted $\gamma_z$ characterises the selected, heterogeneous sample \citep{Bolton2012,Sonnenfeld2013,stz1902}, and $\gamma_0$ is its extrapolated zero-redshift intercept. The inference targets dynamical model sensitivities; cosmological parameters, including $H_0$, are not fitted.

The assumed S\'ersic indices, PSF values, and broader latent-slope widths define sensitivity scenarios in the absence of individual image-fitting and spectroscopic-extraction likelihoods. Within this scope, the likelihood, recovery, and sensitivity comparisons quantify how light-profile assumptions can affect a common-anisotropy preference. Interpreting that preference as an individual orbital property, or attributing the observed residuals to a unique physical cause, requires additional photometric and kinematic constraints.

\section{Conclusions}\label{sec:conclusions}
We quantified the dependence of anisotropy and density-slope inference on tracer-profile assumptions using 107 galaxy-scale lenses and distinct \pantheon\ counterparts. The main findings are:
\begin{enumerate}
\item The baseline individual-slope power-law tracer yields $\Delta\mathrm{BIC}_0=6.74$ and $\Delta\mathrm{BIC}_{0.18}=41.51$ when each specified fixed reference is compared with a common free anisotropy parameter.
\item Hernquist and S\'ersic $n=4$ alternatives remove this criterion advantage and reduce the slope changes induced by fixing beta. The contrasting results quantify sensitivity to the photometric representation, which uses different information in the power-law and curved-tracer cases.
\item The comparison also depends on distance uncertainty, survey composition, and the input slopes. The 95-lens lower-uncertainty subset gives a power-law isotropy difference of 2.51. The principal tracer dependence persists across the finite-matching and alternative-assignment checks, while the strength of the power-law preference varies.
\item Under the specified conditional noise model, fitting a power-law tracer to isotropic S\'ersic $n=4$ and Hernquist simulations produces $\Delta\mathrm{BIC}_0>6$ in 60.0\% and 50.0\% of realisations, respectively. The corresponding matched-tracer frequencies are 0.5\% and 1.0\%. These controlled experiments show how tracer mismatch can generate an effective anisotropy preference.
\end{enumerate}
The combined comparisons motivate joint assessment of anisotropy, light-profile representation, photometric uncertainty, and aperture weighting. They provide a quantitative basis for testing light-profile assumptions before interpreting a fitted common beta as a population orbital property. Consistent imaging constraints and spatially resolved spectroscopy can further separate tracer structure from orbital anisotropy.

\section*{Data availability}\label{sec:data_avail}
The \pantheon\ data and covariance are available from the official release at \url{https://github.com/PantheonPlusSH0ES/DataRelease}. Original lens surveys are cited in Table~\ref{tab:surveys}, and the compiled lens parameters are listed in Table A1 of \citet{stz1902}. The individual luminosity-slope measurements were provided by Y.~Chen. The analysis code, input data, frozen matched catalogue, derived summary tables, and diagnostic records accompanying this manuscript are publicly available at \url{https://github.com/HUJIAN0000/sgl-sn-anisotropy}, in \href{https://github.com/HUJIAN0000/sgl-sn-anisotropy/releases/tag/v1.1.0}{release \texttt{v1.1.0}}. The release includes posterior summaries and trace plots; retained posterior samples and full sampler backends are not included.

\begin{acknowledgements}

We acknowledge the use of the \pantheon\ supernova data and are grateful to
Dr.~Yun~Chen for the individual luminosity density-slope measurements. This work is supported by the National Natural Science Foundation of China (Grant No.~12563004), the Yunnan Provincial Science and Technology Department (Grant No.~202501BA070001-064), the Yunnan Basic Research Projects (Grant No.~202501AT070439), and the Local Universities Joint Special Project -- General Project (Grant No.~202301A0070247) under the Basic Research Program.

\end{acknowledgements}

\appendix
\section{Analytic power-law normalisation}\label{sec:analytic}
With $\xi=\gamma+\delta_i-2$, the dimensionless function in Eq.~(\ref{eq:sigma_th}) is
\begin{equation}
\begin{split}
F={}&\frac{3-\delta_i}{(\xi-2\beta)(3-\xi)}
\left[\frac{\Gamma((\xi-1)/2)}{\Gamma(\xi/2)}-
\beta\frac{\Gamma((\xi+1)/2)}{\Gamma((\xi+2)/2)}\right]\\
&\times\frac{\Gamma(\gamma/2)\Gamma(\delta_i/2)}{\Gamma((\gamma-1)/2)\Gamma((\delta_i-1)/2)}.
\end{split}
\end{equation}
For $\gamma=\delta_i=2$ and $\beta=0$, $F=1/(2\sqrt\pi)$, giving $\sigma^2=c^2\theta_E/(4\pi R)$, the singular-isothermal-sphere normalisation. The numerical power-law engine reproduces this limit and general convergent analytic moments. Formal analytic continuations outside the convergent domain are not used for the main posterior.

\section{Covariance propagation}\label{sec:propagation}
Let $q_i=1-R_i$. The derivatives are
\begin{equation}
\frac{\partial R_i}{\partial\Delta m_i}=-\frac{\ln10}{5}q_i,\qquad
\frac{\partial\sigma_i}{\partial R_i}=-\frac{\sigma_i}{2R_i}.
\end{equation}
The diagonal Jacobians give $\mathbf C_R=\mathbf J_{Rm}\mathbf C_{\Delta m}\mathbf J_{Rm}^{T}$ and $\mathbf C_{\sigma,{\rm SN}}=\mathbf J_{\sigma R}\mathbf C_R\mathbf J_{\sigma R}^{T}$. The latter is recalculated at every likelihood evaluation. This first-order construction retains covariance but does not represent finite-redshift matching errors or a nonlinear distance posterior.

For a fiducial-assisted redshift correction, define $q_i^{\rm match,fid}$ by applying the matching formula to noiseless model magnitudes at the actual counterpart redshifts and $q_i^{\rm target,fid}=\chi_{l,i}/\chi_{s,i}$. With $a_i=q_i^{\rm target,fid}/q_i^{\rm match,fid}$, the check uses $R'_i=1-a_i(1-R_i)$ and $(C'_R)_{ij}=a_i(C_R)_{ij}a_j$. The assumed background dependence is explicit.

\section{Numerical verification and sampling diagnostics}\label{sec:numerical}
The radial calculations are checked against doubled radial resolution, a factor-ten extension of the radial limits, and doubled Abel/angular nodes. Each curved-tracer case is evaluated at its three likelihood maxima and 50 independent feasible perturbations near the free maximum. We require $|\Delta\ln\mathcal L|<0.1$ and maximum relative dispersion changes below $5\times10^{-4}$. Differences smaller than the numerical uncertainty are interpreted as effectively equal likelihoods.

To accelerate sampling, a tensor Chebyshev representation of $\log(\sigma^2 R/A(\gamma))$ is built from direct integrations, where $A(\gamma)=2\Gamma(\gamma/2)/[\sqrt\pi\Gamma((\gamma-1)/2)]$ is evaluated analytically. Removing this known factor improves behaviour near the lower gamma boundary. The representation is checked using 120 independent off-grid parameter sets across the supported gamma and beta range, requiring relative dispersion error below $2\times10^{-6}$, plus explicit likelihood checks against the direct engine. The cache includes the input, profile, grid and implementation hashes.

Posterior checks include rank-normalised and folded split $\widehat R<1.01$ \citep{Vehtari2021}, autocorrelation-based effective sample size (ESS) at least 2000, retained chain lengths exceeding $100\tau$, and changes in estimated $\tau$ below 5\%. Separate ensembles supply independent runs; interacting walkers are not declared independent chains. Chain lengths, diagnostic arrays and trace plots are retained in the reproduction material. The same convergence thresholds are applied throughout the sampling.
All 16 main and eight latent-slope posterior calculations pass the stated diagnostics. The largest reported rank-normalised split $\widehat R$ is 1.0080, and the smallest approximate ESS is 25868. The curved-tracer numerical audit additionally evaluates 24 retained posterior draws per tracer/prior combination, 192 points in total, against doubled quadrature resolution. These checks supplement the likelihood-maximum and off-grid interpolation tests.

\section{Sensitivity comparisons}\label{sec:sensitivity}
Table~\ref{tab:sensitivity} compares each beta hypothesis within its own stated sample and scenario. Absolute BIC values are not compared between different data memberships. Coherent delta offsets are assumed perturbations; mismatch corrections use stated fiducial backgrounds. The separate P3 control uses the power-law tracer and one additional structural coefficient.
\begin{table*}[t]\centering\small
\caption{Sensitivity comparisons. Positive fixed-minus-free differences favour the free model within the given row. Subset differences conflate information loss, selection and changed design; they are not estimates of a causal survey effect.}
\label{tab:sensitivity}\begin{tabular}{llrrr}
\toprule
Scenario & Tracer & $N$ & $\Delta\mathrm{BIC}_0$ & $\Delta\mathrm{BIC}_{0.18}$ \\
\midrule
$\sigma_R/R\le0.20$ & Power law & 95 & 2.51 & 29.90 \\
$\sigma_R/R\le0.20$ & Hernquist & 95 & -4.04 & -2.86 \\
$\sigma_R/R\le0.20$ & S\'ersic $n=4$ & 95 & -4.54 & -4.40 \\
\addlinespace[3pt]
Exclude BELLS family & Power law & 89 & 7.40 & 38.86 \\
Exclude BELLS family & Hernquist & 89 & -2.62 & -0.60 \\
Exclude BELLS family & S\'ersic $n=4$ & 89 & -4.09 & -3.21 \\
\addlinespace[3pt]
Exclude S4TM & Power law & 71 & -1.82 & 9.69 \\
Exclude S4TM & Hernquist & 71 & -4.10 & -3.54 \\
Exclude S4TM & S\'ersic $n=4$ & 71 & -4.17 & -4.26 \\
\addlinespace[3pt]
Fiducial mismatch ($\Omega_m=0.25$) & Power law & 107 & 6.68 & 41.22 \\
Fiducial mismatch ($\Omega_m=0.25$) & Hernquist & 107 & -3.84 & -2.17 \\
Fiducial mismatch ($\Omega_m=0.25$) & S\'ersic $n=4$ & 107 & -4.67 & -4.26 \\
\addlinespace[3pt]
Fiducial mismatch ($\Omega_m=0.30$) & Power law & 107 & 6.68 & 41.23 \\
Fiducial mismatch ($\Omega_m=0.30$) & Hernquist & 107 & -3.84 & -2.17 \\
Fiducial mismatch ($\Omega_m=0.30$) & S\'ersic $n=4$ & 107 & -4.67 & -4.26 \\
\addlinespace[3pt]
Fiducial mismatch ($\Omega_m=0.35$) & Power law & 107 & 6.68 & 41.23 \\
Fiducial mismatch ($\Omega_m=0.35$) & Hernquist & 107 & -3.84 & -2.17 \\
Fiducial mismatch ($\Omega_m=0.35$) & S\'ersic $n=4$ & 107 & -4.67 & -4.26 \\
\addlinespace[3pt]
Heliocentric transfer ($\Omega_m=0.30$) & Power law & 107 & 6.64 & 41.17 \\
Heliocentric transfer ($\Omega_m=0.30$) & Hernquist & 107 & -3.85 & -2.19 \\
Heliocentric transfer ($\Omega_m=0.30$) & S\'ersic $n=4$ & 107 & -4.67 & -4.27 \\
\addlinespace[3pt]
Slope offset $\Delta\delta=-0.10$ & Power law & 107 & 33.10 & 97.98 \\
\addlinespace[3pt]
Slope offset $\Delta\delta=-0.05$ & Power law & 107 & 16.09 & 69.47 \\
\addlinespace[3pt]
Slope offset $\Delta\delta=+0.05$ & Power law & 107 & 2.21 & 21.41 \\
\addlinespace[3pt]
Slope offset $\Delta\delta=+0.10$ & Power law & 107 & 0.06 & 10.58 \\
\bottomrule
\end{tabular}
\end{table*}

\section{Sampling convergence diagnostics and matching variants}\label{sec:repro}
For each main posterior we use four independently initialised ensembles; each latent-slope posterior uses four independent chains. Main ensembles contain 15000 or 25000 steps with 3000 discarded, and latent-slope chains contain 80000--320000 steps with 5000 discarded. Depending on the beta treatment and common-offset scenario, the latent analyses have 110--112 varying coordinates. Table~\ref{tab:diagnostics} lists the accepted lengths, maximum rank-normalised split $\widehat R$, and minimum autocorrelation-based approximate ESS. The latent diagnostics include all varying slope coordinates and the common offset where fitted.

We retain the full sampler checkpoints locally. The public release contains the matched supernova row indices, covariance blocks, configurations, diagnostic summaries, and trace plots. The magnitude covariance is symmetrised as $(C+C^T)/2$ to remove the published rounding-level asymmetry (maximum $3\times10^{-8}$). Table~\ref{tab:matching} gives the alternative assignments, and Table~\ref{tab:p3posterior} reports the P3 posterior parameters. Approximate ESS is used alongside chain-length, autocorrelation-stability, between-run, and trace checks.
\begin{table*}[t]\centering\small\caption{Accepted posterior diagnostics. Each row contains four independent ensembles or chains. The minimum approximate ESS and maximum rank-normalised split $\widehat R$ include every varying latent coordinate for the slope-error products.}\label{tab:diagnostics}\begin{tabular}{lllrrr}
\toprule
Analysis & Model or slope-error scenario & $\beta$ treatment & Steps & $\max\widehat R$ & Min. approx. ESS \\
\midrule
Main & Power law & Free, flat & 25000 & 1.0057 & 44560 \\
Main & Power law & Free, Gaussian & 25000 & 1.0055 & 56369 \\
Main & Power law & Fixed 0 & 25000 & 1.0045 & 68024 \\
Main & Power law & Fixed 0.18 & 25000 & 1.0032 & 65834 \\
\addlinespace[3pt]
Main & Hernquist & Free, flat & 25000 & 1.0079 & 47316 \\
Main & Hernquist & Free, Gaussian & 25000 & 1.0041 & 57722 \\
Main & Hernquist & Fixed 0 & 25000 & 1.0026 & 71119 \\
Main & Hernquist & Fixed 0.18 & 15000 & 1.0080 & 38886 \\
\addlinespace[3pt]
Main & S\'ersic $n=4$ & Free, flat & 25000 & 1.0064 & 41319 \\
Main & S\'ersic $n=4$ & Free, Gaussian & 25000 & 1.0041 & 56329 \\
Main & S\'ersic $n=4$ & Fixed 0 & 15000 & 1.0079 & 38693 \\
Main & S\'ersic $n=4$ & Fixed 0.18 & 15000 & 1.0045 & 39428 \\
\addlinespace[3pt]
Main & P3 power law & Free, flat & 25000 & 1.0069 & 34611 \\
Main & P3 power law & Free, Gaussian & 25000 & 1.0047 & 45247 \\
Main & P3 power law & Fixed 0 & 25000 & 1.0037 & 53122 \\
Main & P3 power law & Fixed 0.18 & 25000 & 1.0044 & 50849 \\
\addlinespace[3pt]
Latent slopes & $\sigma_{\delta,i}=0.01\widehat\delta_i$ & Free, flat & 80000 & 1.0001 & 50518 \\
Latent slopes & $\sigma_{\delta,i}=0.01\widehat\delta_i$ & Fixed 0 & 80000 & 1.0001 & 72233 \\
\addlinespace[3pt]
Latent slopes & $\sigma_\delta=0.05$ & Free, flat & 80000 & 1.0001 & 43017 \\
Latent slopes & $\sigma_\delta=0.05$ & Fixed 0 & 120000 & 1.0001 & 73788 \\
\addlinespace[3pt]
Latent slopes & $\sigma_\delta=0.10$ & Free, flat & 120000 & 1.0001 & 39199 \\
Latent slopes & $\sigma_\delta=0.10$ & Fixed 0 & 320000 & 1.0001 & 67473 \\
\addlinespace[3pt]
Latent slopes & $\sigma_\delta=0.05$, $\sigma_b=0.05$ & Free, flat & 80000 & 1.0001 & 42228 \\
Latent slopes & $\sigma_\delta=0.05$, $\sigma_b=0.05$ & Fixed 0 & 80000 & 1.0002 & 25868 \\
\bottomrule
\end{tabular}
\end{table*}
\begin{table*}[t]\centering\small\caption{Alternative matching assignments. Comparisons are within each row; shared triplets use original lens and SN identifiers.}\label{tab:matching}\begin{tabular}{llrrrr}
\toprule
Assignment & Tracer & $N$ & Shared triplets & $\Delta\mathrm{BIC}_0$ & $\Delta\mathrm{BIC}_{0.18}$ \\
\midrule
Tolerance 0.03 & Power law & 104 & 92 & 3.69 & 35.32 \\
Tolerance 0.03 & Hernquist & 104 & 92 & -4.32 & -3.12 \\
Tolerance 0.03 & S\'ersic $n=4$ & 104 & 92 & -4.56 & -4.56 \\
Tolerance 0.07 & Power law & 111 & 81 & 6.90 & 41.99 \\
Tolerance 0.07 & Hernquist & 111 & 81 & -4.32 & -3.04 \\
Tolerance 0.07 & S\'ersic $n=4$ & 111 & 81 & -4.68 & -4.56 \\
Reversed SN order & Power law & 107 & 78 & 6.51 & 41.67 \\
Reversed SN order & Hernquist & 107 & 78 & -3.75 & -2.00 \\
Reversed SN order & S\'ersic $n=4$ & 107 & 78 & -4.66 & -4.22 \\
\bottomrule
\end{tabular}
\end{table*}
\begin{table*}[t]\centering\small\caption{P3 power-law posterior medians and central 68\% intervals, at the sample median redshift and median structural predictor.}\label{tab:p3posterior}\begin{tabular}{lrrrrr}
\toprule
$\beta$ treatment & $\gamma_p$ & $\gamma_0$ & $\gamma_z$ & $\gamma_s$ & $f_{\rm int}$ \\
\midrule
Free, flat & $2.014_{-0.069}^{+0.050}$ & $2.142_{-0.112}^{+0.101}$ & $-0.841_{-0.276}^{+0.262}$ & $-0.042_{-0.090}^{+0.091}$ & $0.083_{-0.011}^{+0.012}$ \\
Free, Gaussian & $1.873_{-0.056}^{+0.051}$ & $2.038_{-0.168}^{+0.145}$ & $-0.947_{-0.312}^{+0.348}$ & $-0.027_{-0.125}^{+0.117}$ & $0.086_{-0.011}^{+0.012}$ \\
Fixed 0 & $1.810_{-0.050}^{+0.046}$ & $1.912_{-0.182}^{+0.166}$ & $-0.845_{-0.343}^{+0.387}$ & $-0.074_{-0.140}^{+0.132}$ & $0.089_{-0.011}^{+0.012}$ \\
Fixed 0.18 & $1.731_{-0.057}^{+0.055}$ & $1.510_{-0.172}^{+0.186}$ & $-0.310_{-0.386}^{+0.399}$ & $-0.333_{-0.107}^{+0.129}$ & $0.108_{-0.012}^{+0.013}$ \\
\bottomrule
\end{tabular}
\end{table*}

\section{Joint posterior distributions}\label{sec:joint_posteriors}
The following triangles show all fitted global parameters for each of the four anisotropy treatments. Every distribution is computed from the retained samples for the primary 107-lens catalogue. The black dashed line on each diagonal marks the full-sample median; the annotations give the median and central 68\% interval. The sienna and tan bands show the 16th--84th and 2.275th--97.725th percentile intervals, respectively, labelled $1\sigma$ and $2\sigma$ for display. The inner and outer blue contours enclose 68\% and 95\% of each two-dimensional marginal density; these joint probabilities are distinct from the diagonal equal-tailed intervals. GetDist provides the smoothed densities with relative smoothing scales 0.5 in one dimension and 0.7 in two dimensions. All retained draws are used without random subsampling. The density estimator is supplied with the actual hard prior ranges to avoid smoothing into excluded parameter values. Smoothing is a display operation and does not enter the reported numerical intervals.

\begin{figure*}[p]\centering
\includegraphics[width=.49\textwidth]{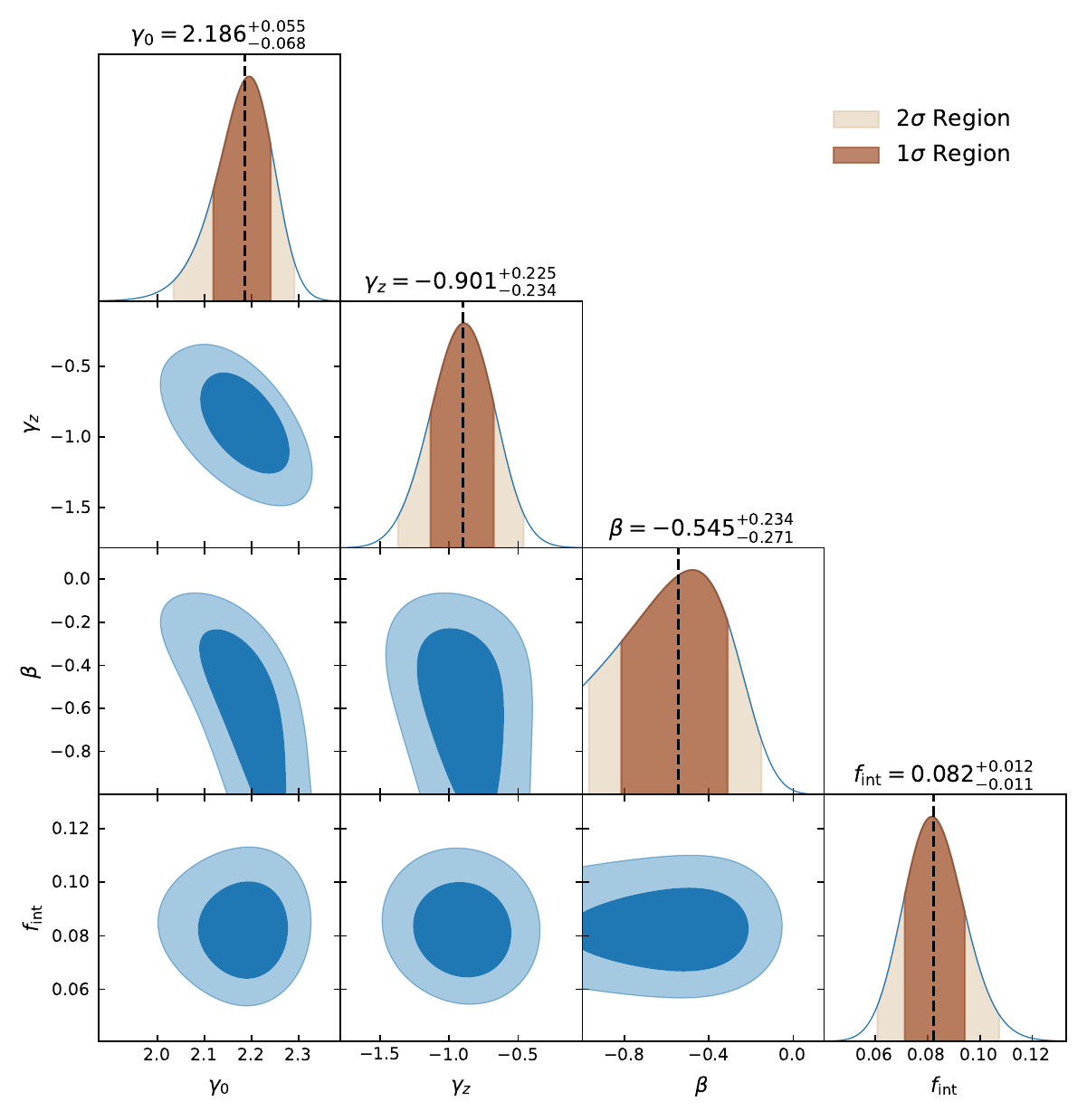}
\includegraphics[width=.49\textwidth]{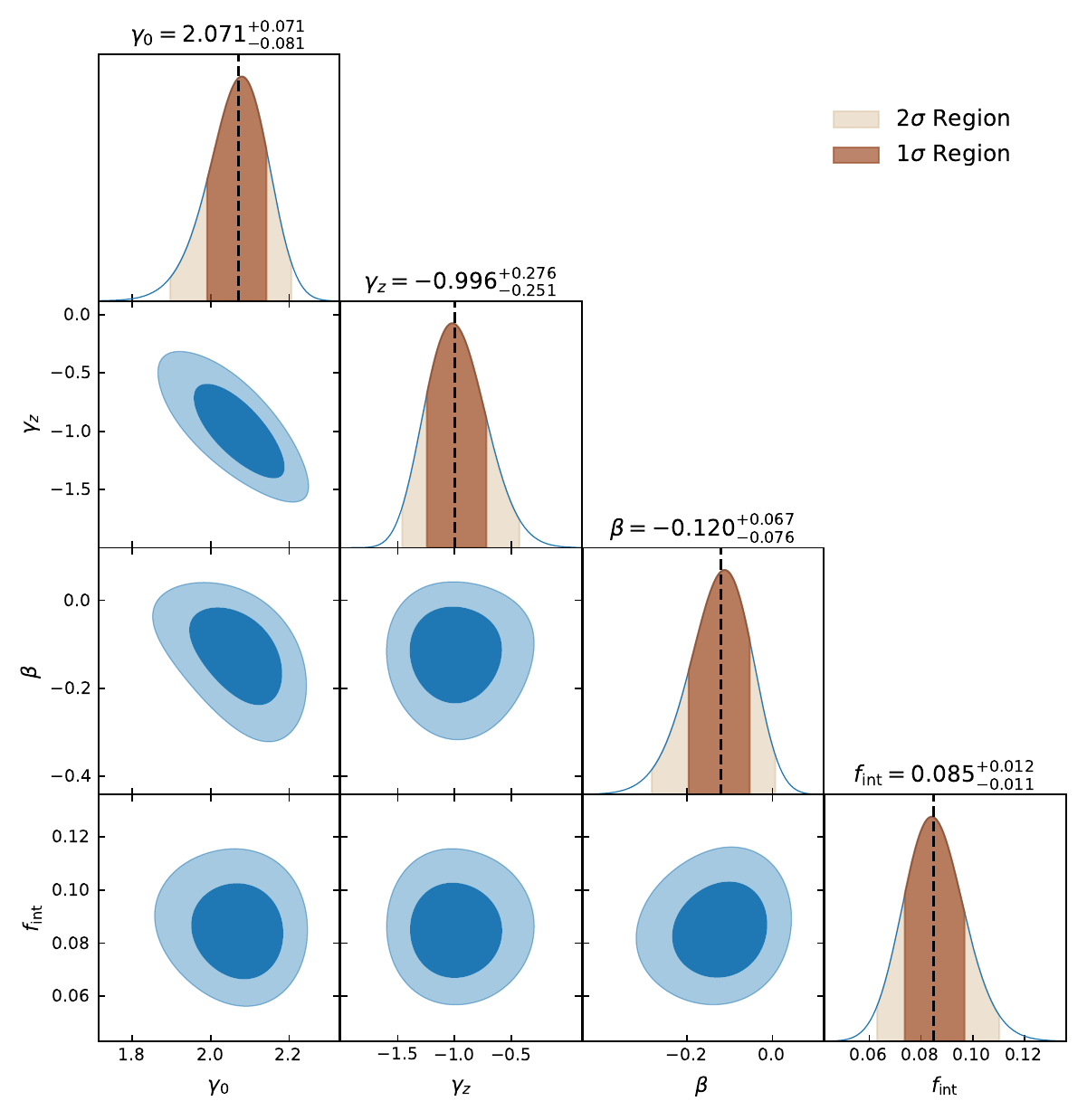}
\caption{P2 individual-slope power-law tracer. Free beta with a uniform prior (left) and the truncated Gaussian prior (right). The marginal and contour conventions are those of Appendix~\ref{sec:joint_posteriors}.}\label{fig:corner_p2_powerlaw_free}\end{figure*}

\begin{figure*}[p]\centering
\includegraphics[width=.49\textwidth]{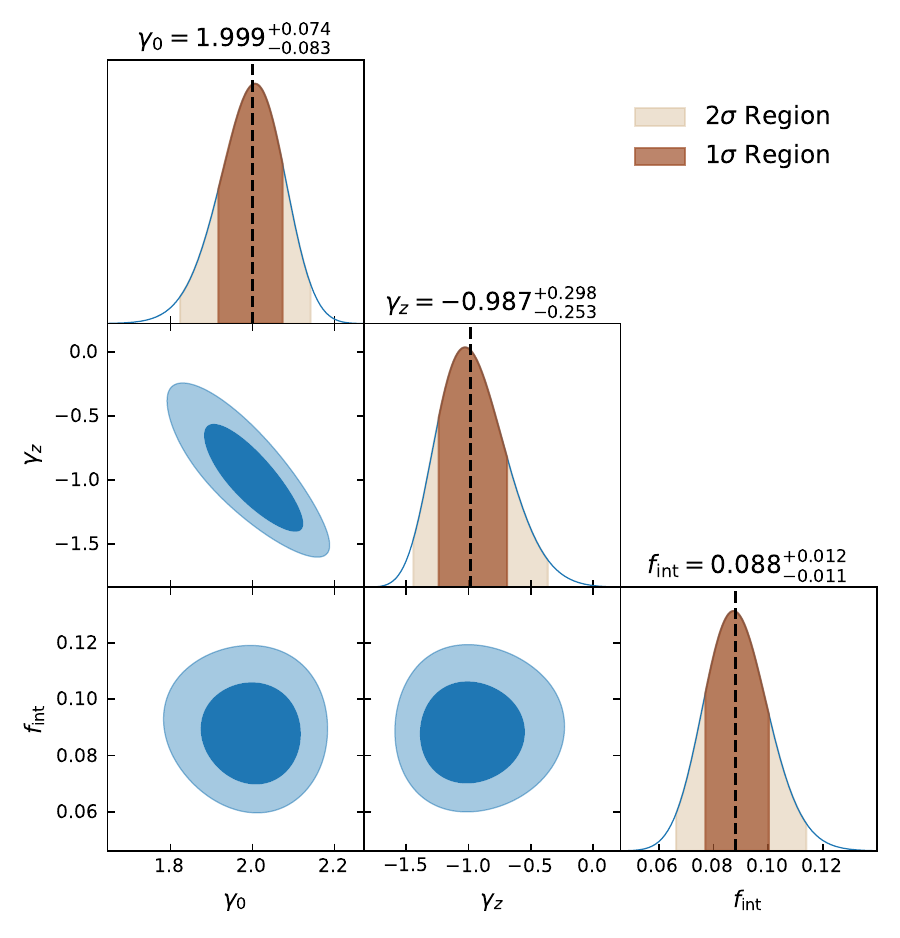}
\includegraphics[width=.49\textwidth]{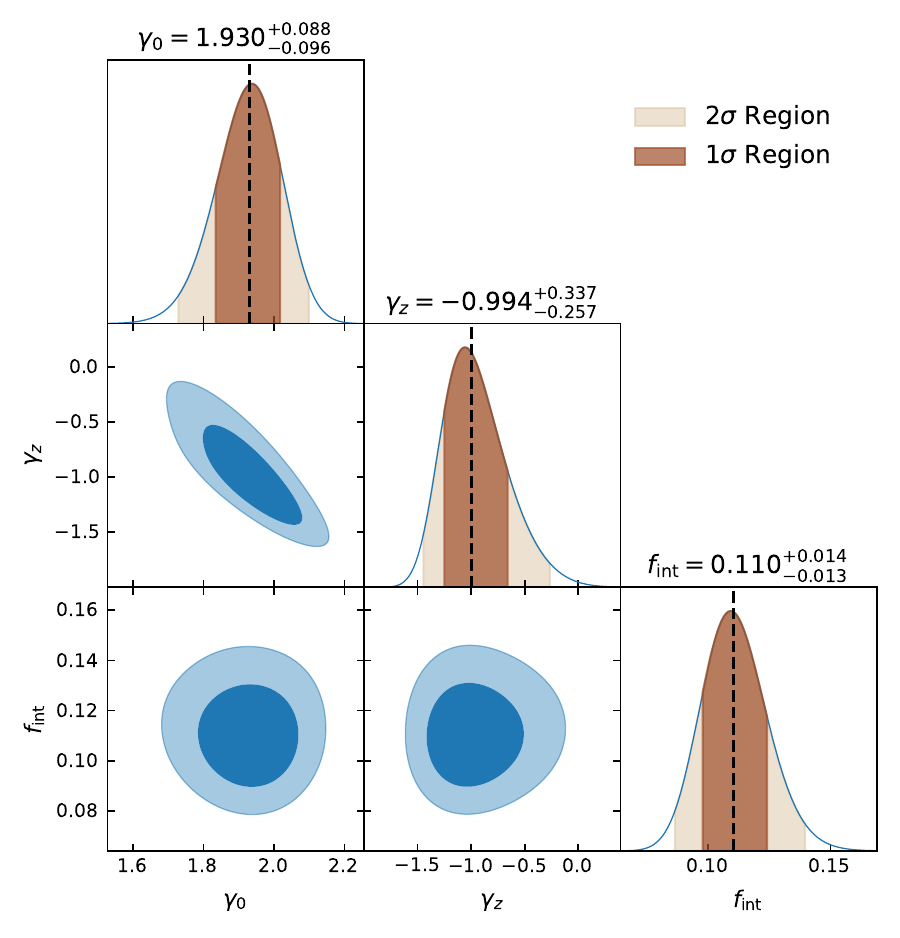}
\caption{P2 individual-slope power-law tracer. Fixed $\beta=0$ (left) and $\beta=0.18$ (right). The marginal and contour conventions are those of Appendix~\ref{sec:joint_posteriors}.}\label{fig:corner_p2_powerlaw_fixed}\end{figure*}

\begin{figure*}[p]\centering
\includegraphics[width=.49\textwidth]{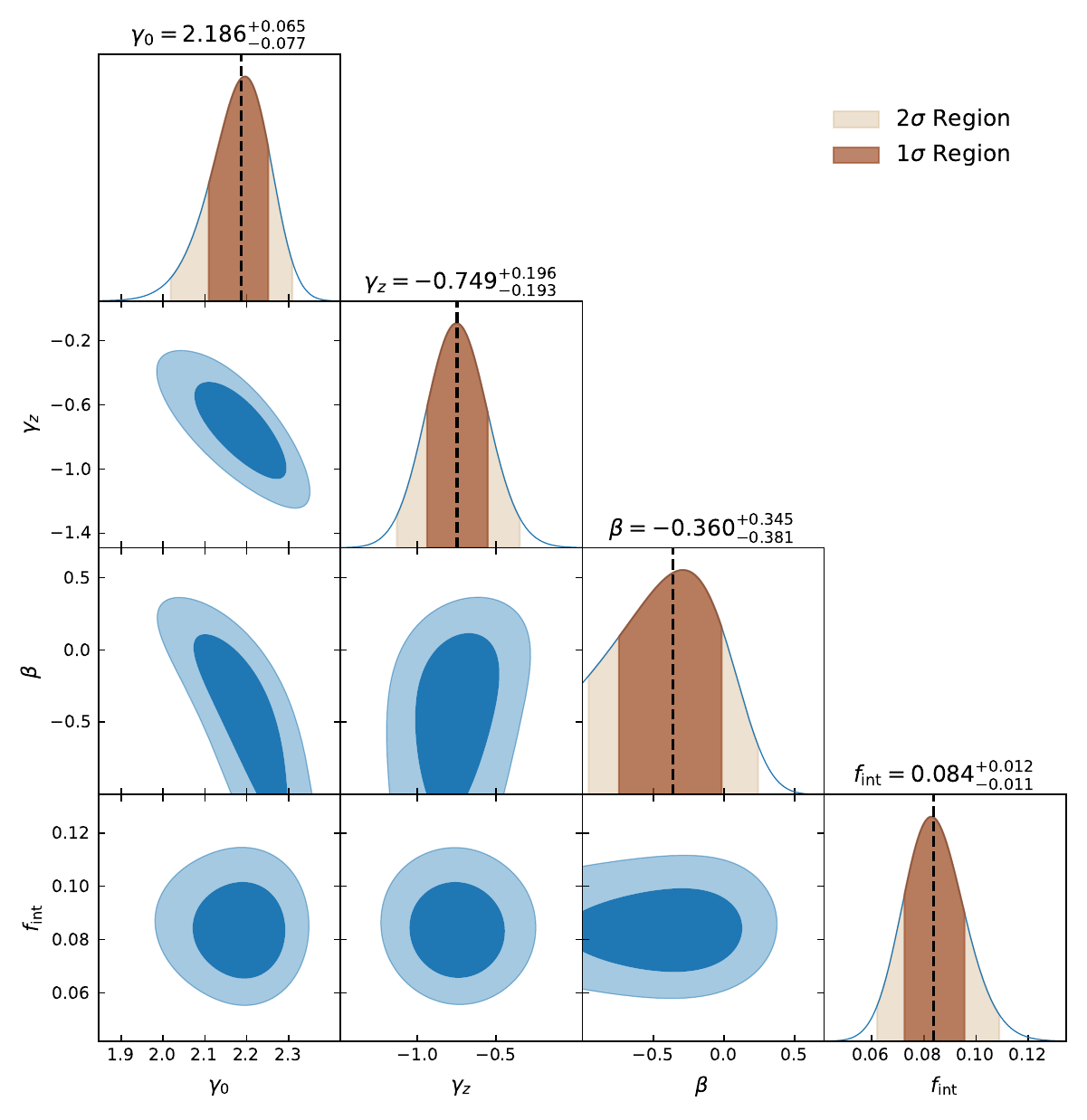}
\includegraphics[width=.49\textwidth]{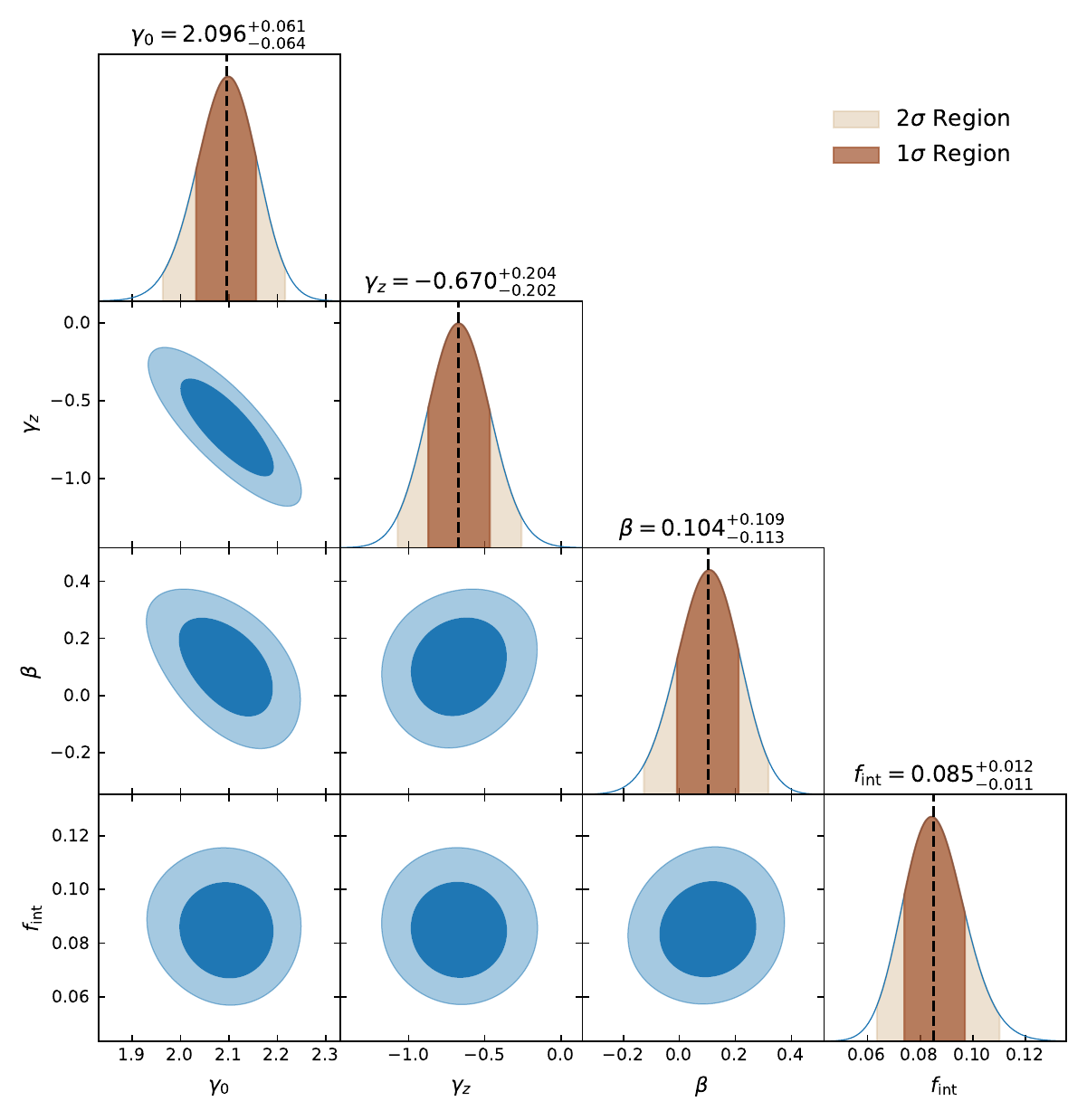}
\caption{P2 Hernquist tracer. Free beta with a uniform prior (left) and the truncated Gaussian prior (right). The marginal and contour conventions are those of Appendix~\ref{sec:joint_posteriors}.}\label{fig:corner_p2_hernquist_free}\end{figure*}

\begin{figure*}[p]\centering
\includegraphics[width=.49\textwidth]{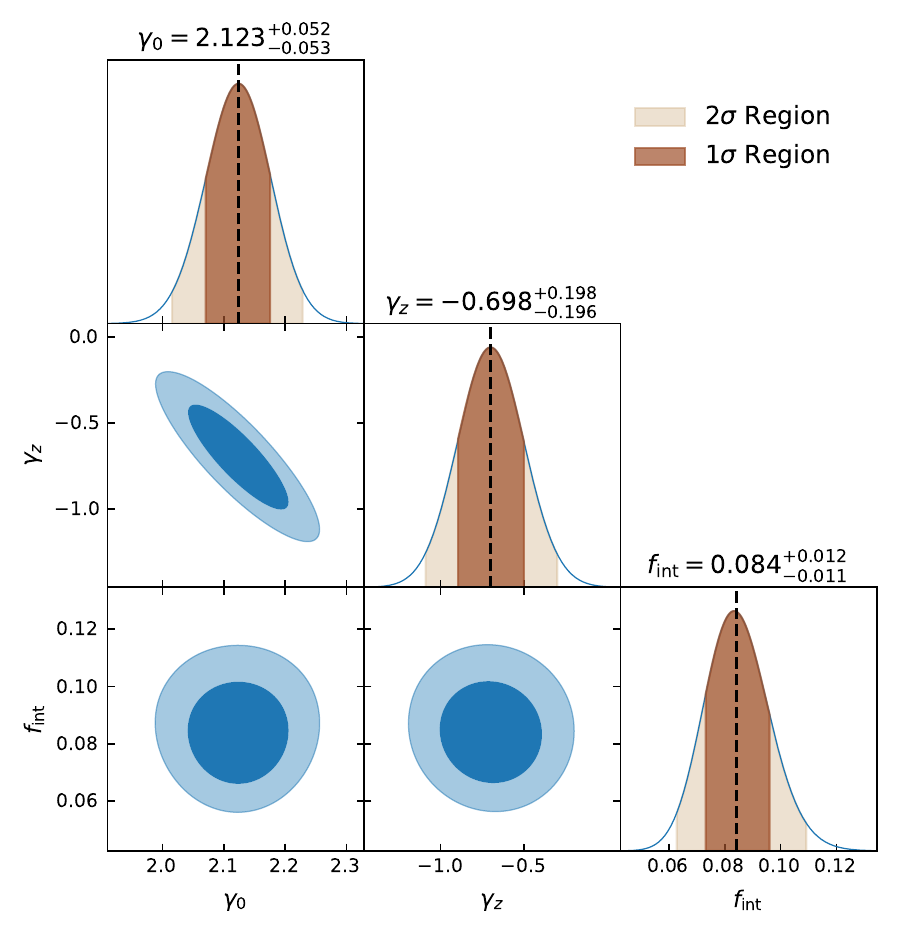}
\includegraphics[width=.49\textwidth]{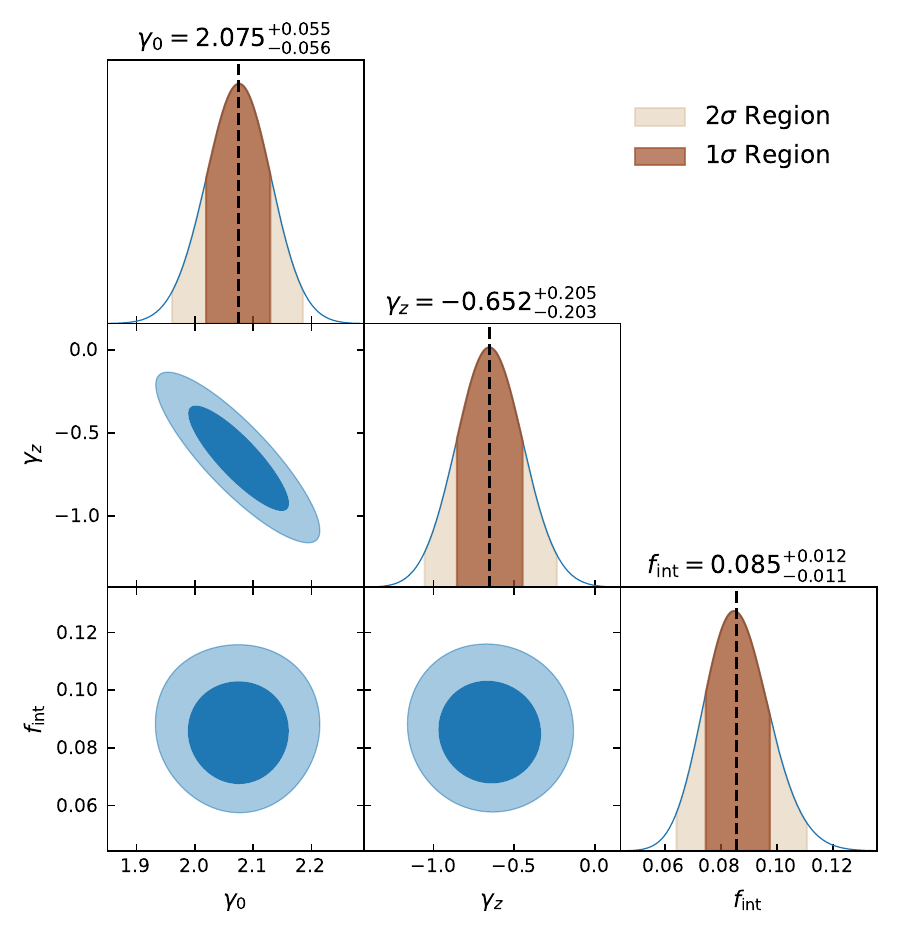}
\caption{P2 Hernquist tracer. Fixed $\beta=0$ (left) and $\beta=0.18$ (right). The marginal and contour conventions are those of Appendix~\ref{sec:joint_posteriors}.}\label{fig:corner_p2_hernquist_fixed}\end{figure*}

\begin{figure*}[p]\centering
\includegraphics[width=.49\textwidth]{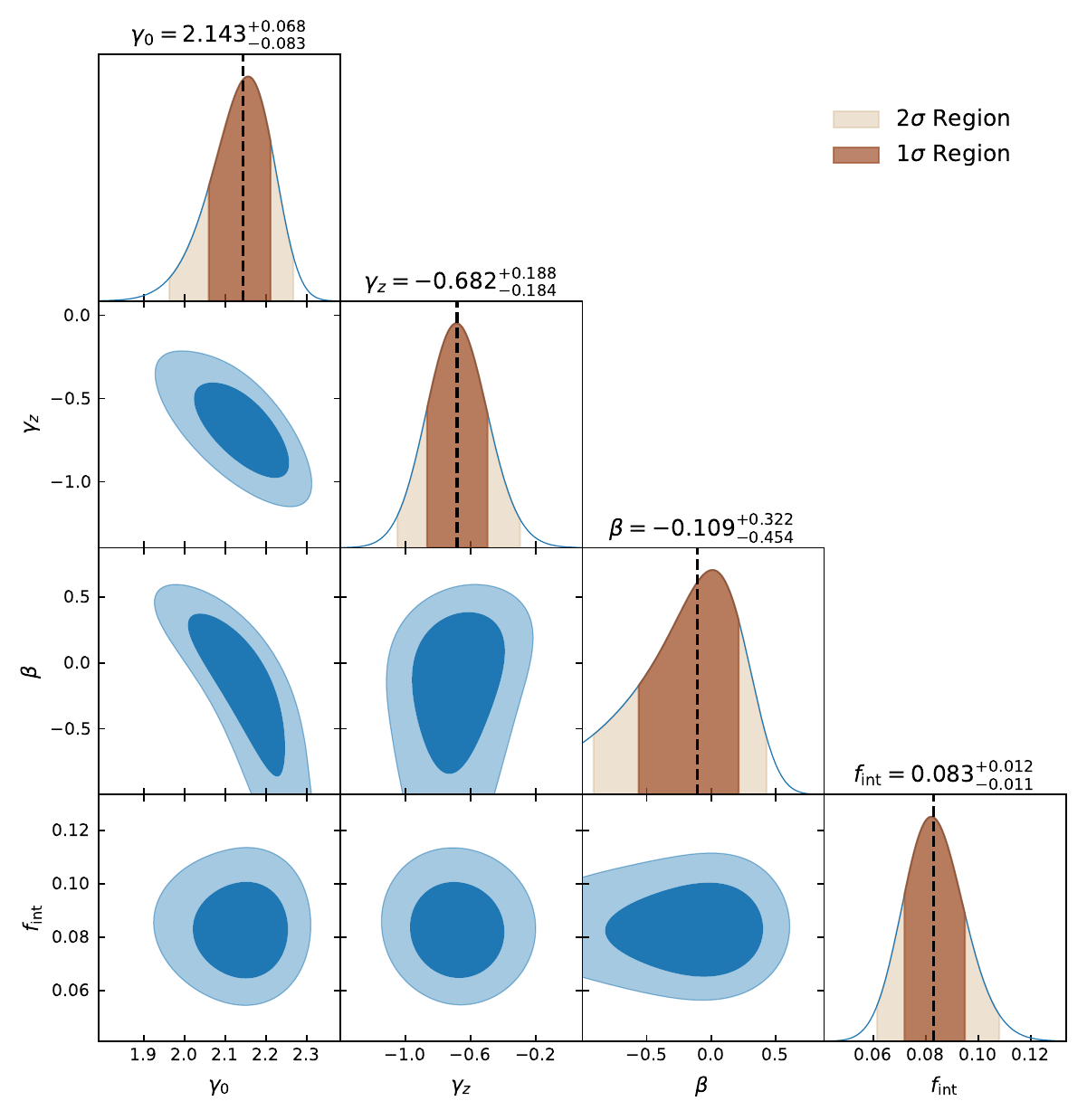}
\includegraphics[width=.49\textwidth]{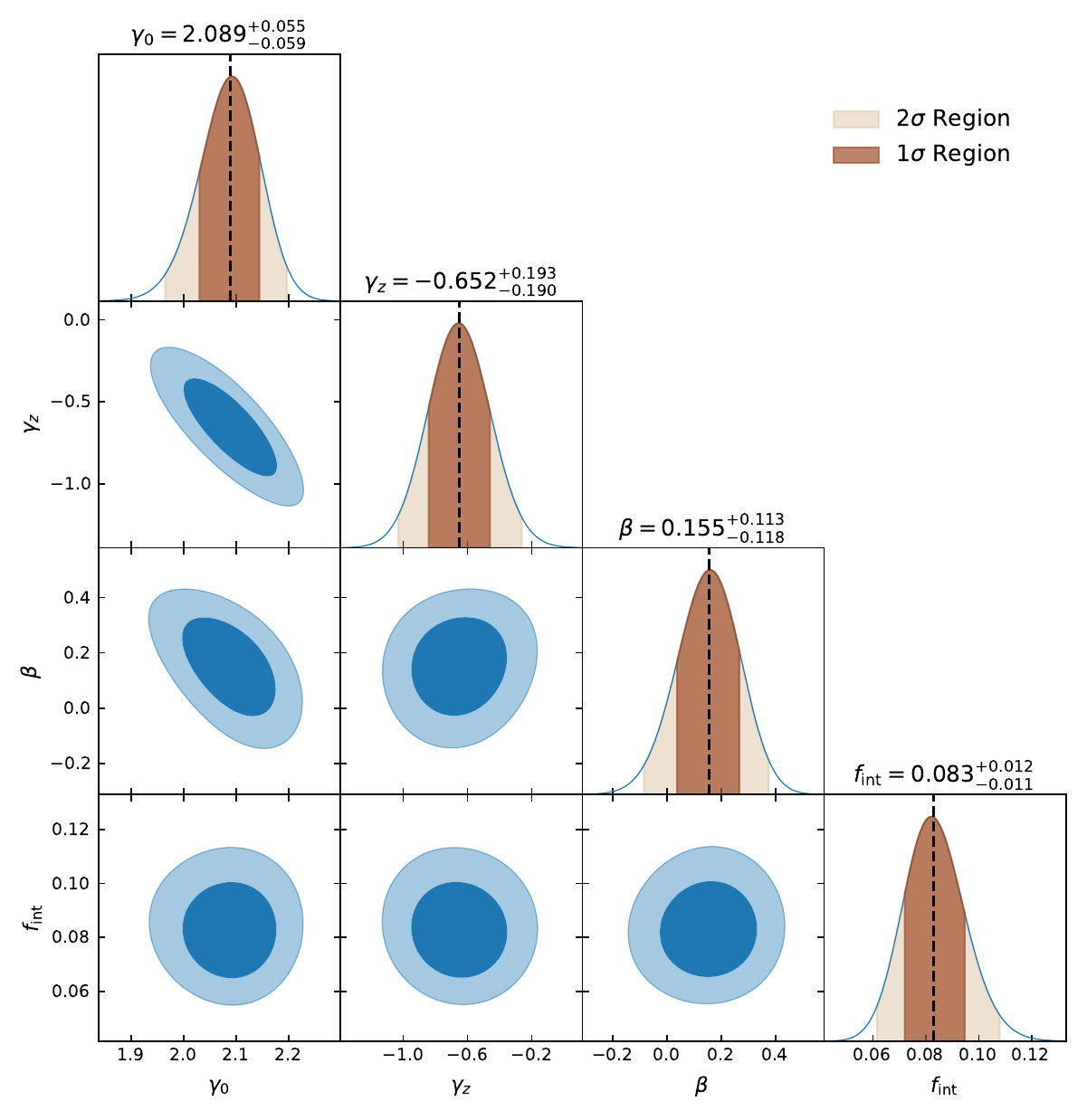}
\caption{P2 S\'ersic $n=4$ tracer. Free beta with a uniform prior (left) and the truncated Gaussian prior (right). The marginal and contour conventions are those of Appendix~\ref{sec:joint_posteriors}.}\label{fig:corner_p2_sersic_free}\end{figure*}

\begin{figure*}[p]\centering
\includegraphics[width=.49\textwidth]{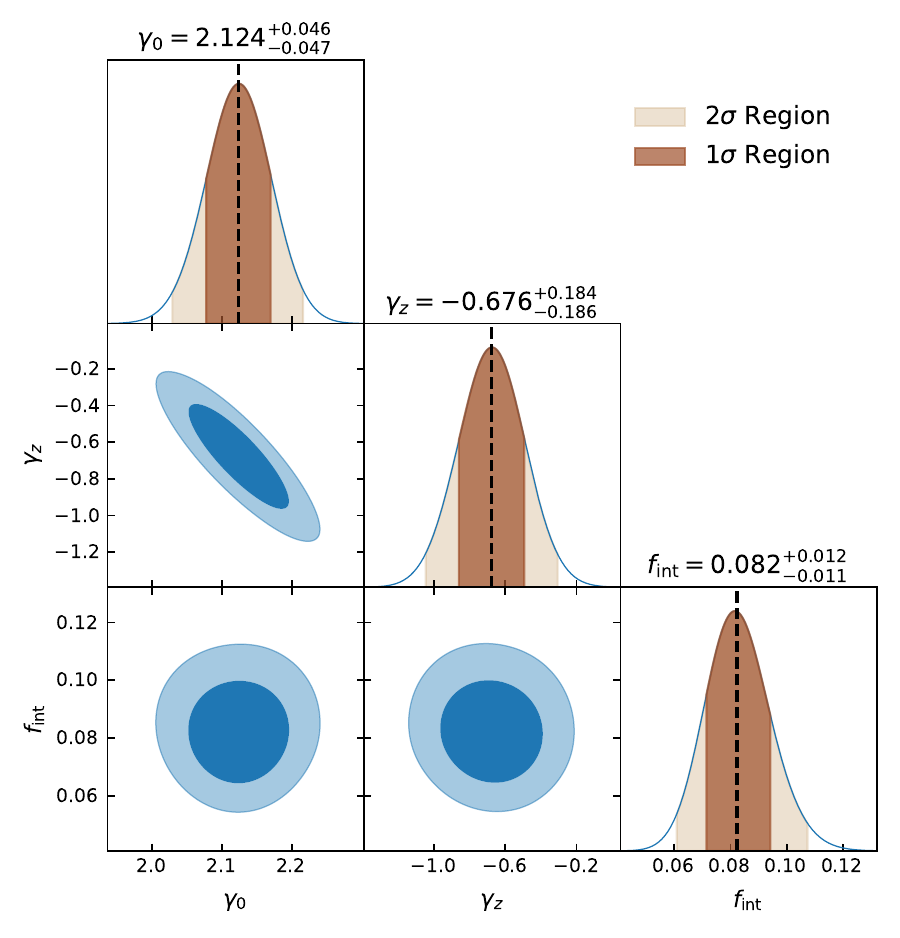}
\includegraphics[width=.49\textwidth]{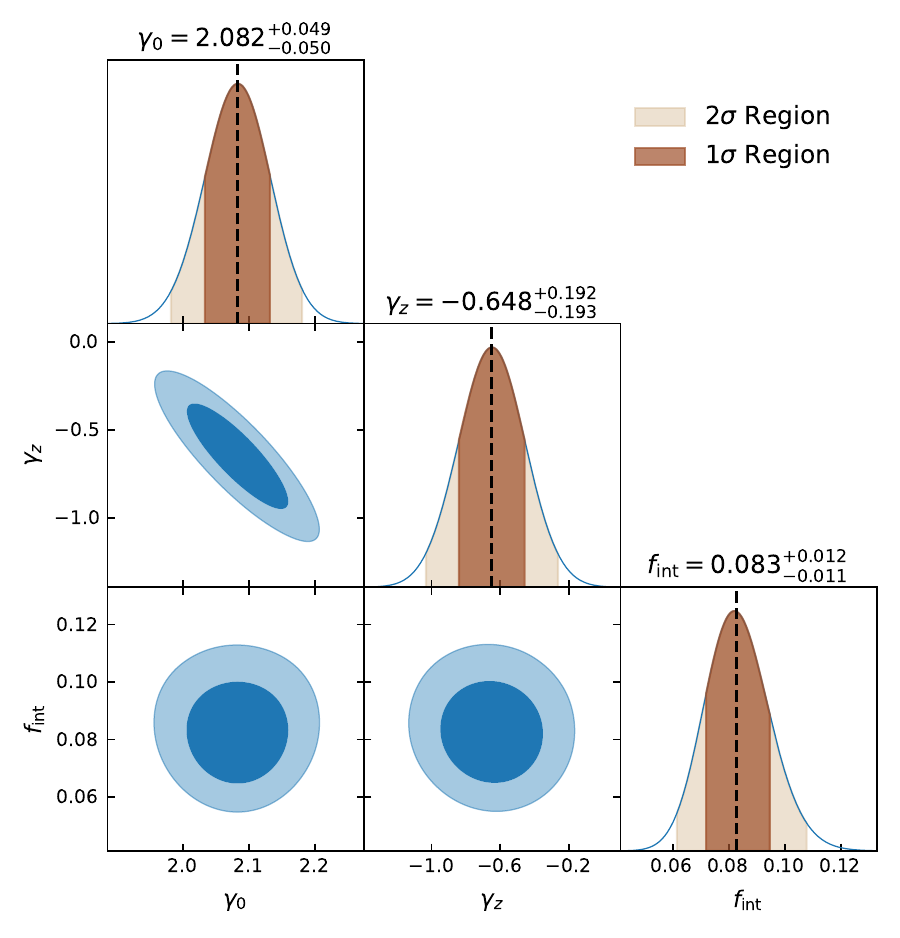}
\caption{P2 S\'ersic $n=4$ tracer. Fixed $\beta=0$ (left) and $\beta=0.18$ (right). The marginal and contour conventions are those of Appendix~\ref{sec:joint_posteriors}.}\label{fig:corner_p2_sersic_fixed}\end{figure*}

\begin{figure*}[p]\centering\includegraphics[width=.86\textwidth]{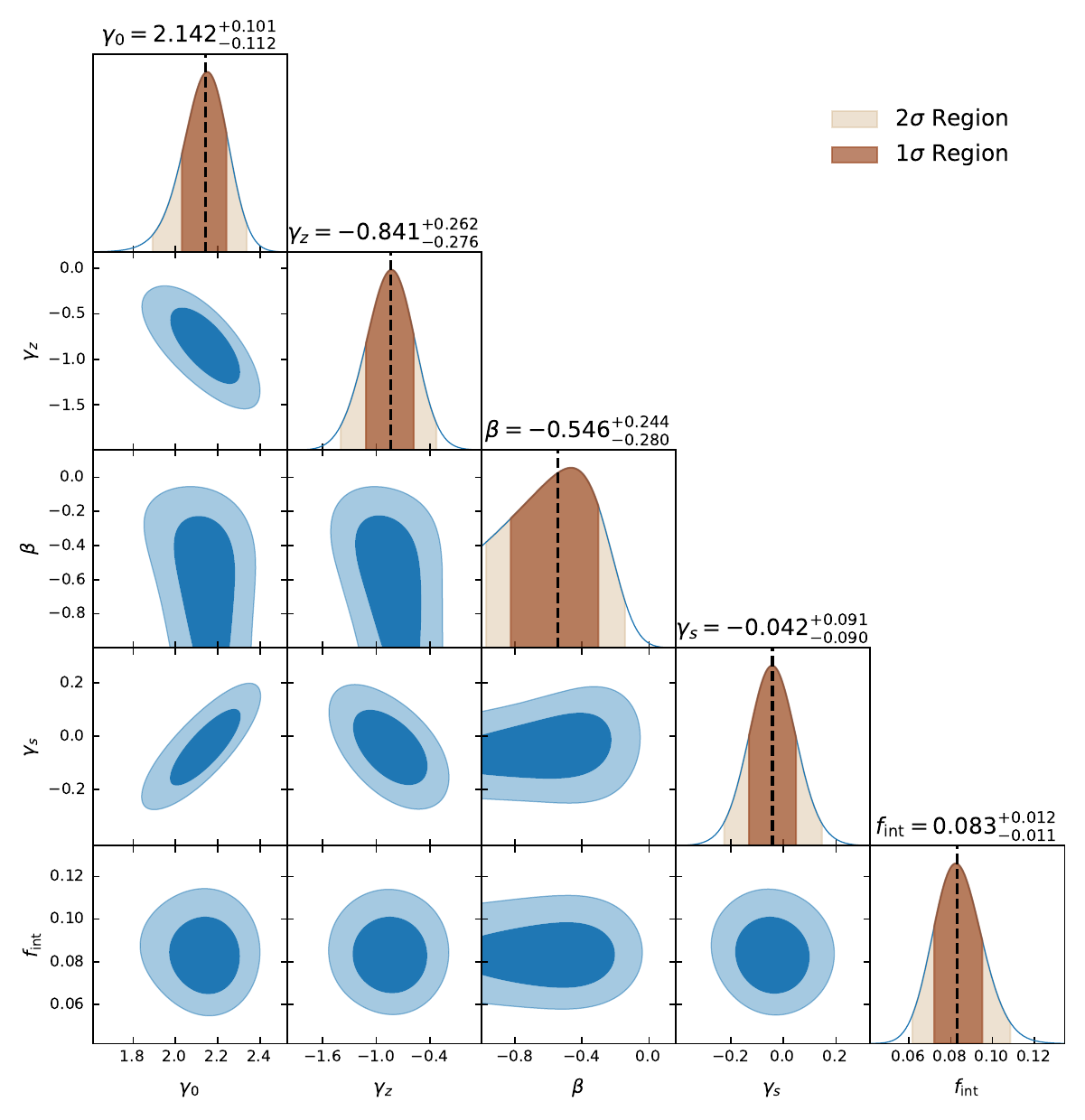}
\caption{P3 individual-slope power-law posterior with a uniform beta prior. The additional structural coefficient $\gamma_s$ multiplies the lensing-based predictor $\log_{10}\widetilde\Sigma_L$.}\label{fig:p3_flat}\end{figure*}

\begin{figure*}[p]\centering\includegraphics[width=.86\textwidth]{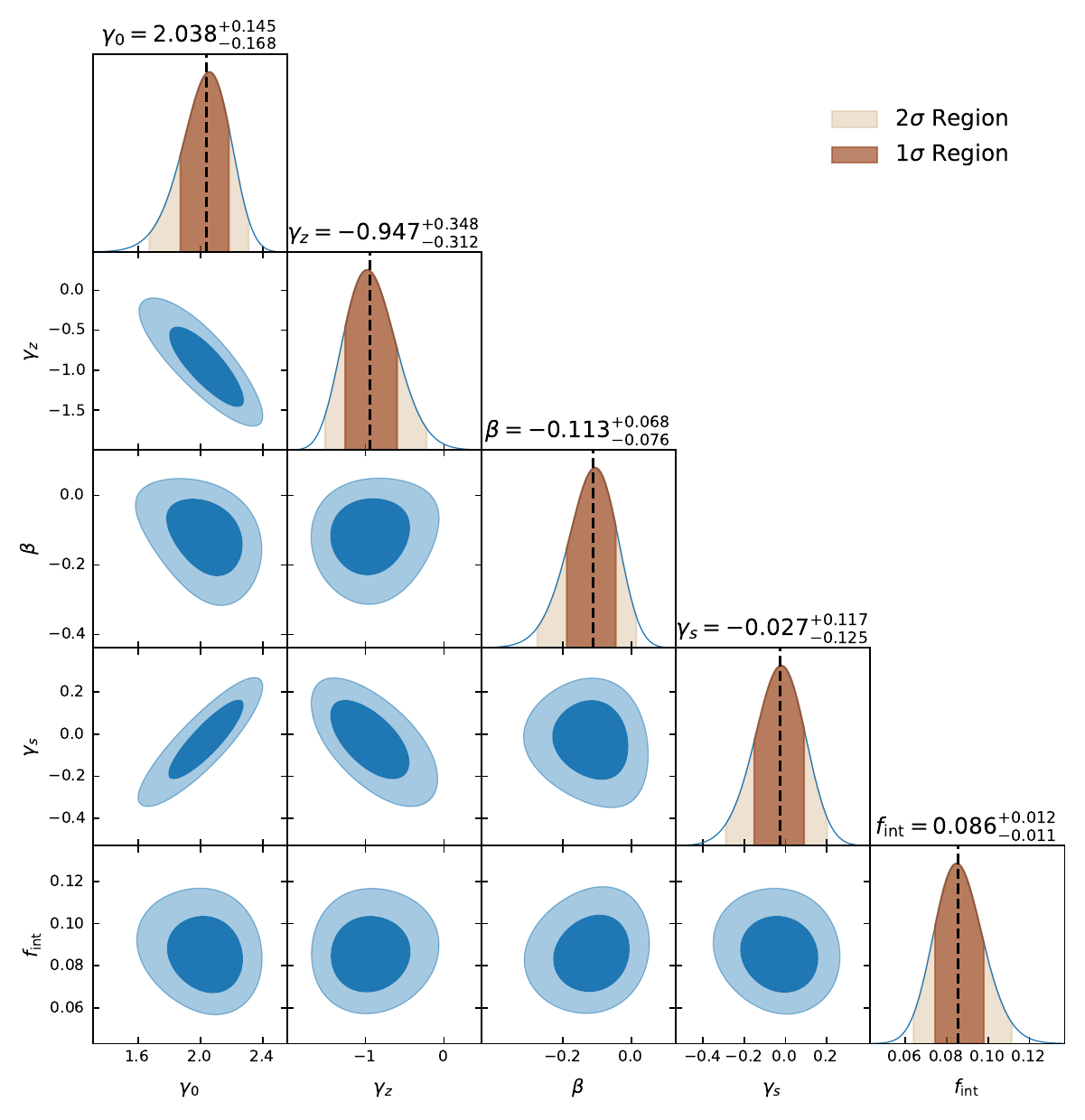}
\caption{P3 individual-slope power-law posterior with the truncated Gaussian beta prior. The additional structural coefficient $\gamma_s$ multiplies the lensing-based predictor $\log_{10}\widetilde\Sigma_L$.}\label{fig:p3_gauss}\end{figure*}

\begin{figure*}[p]\centering
\includegraphics[width=.49\textwidth]{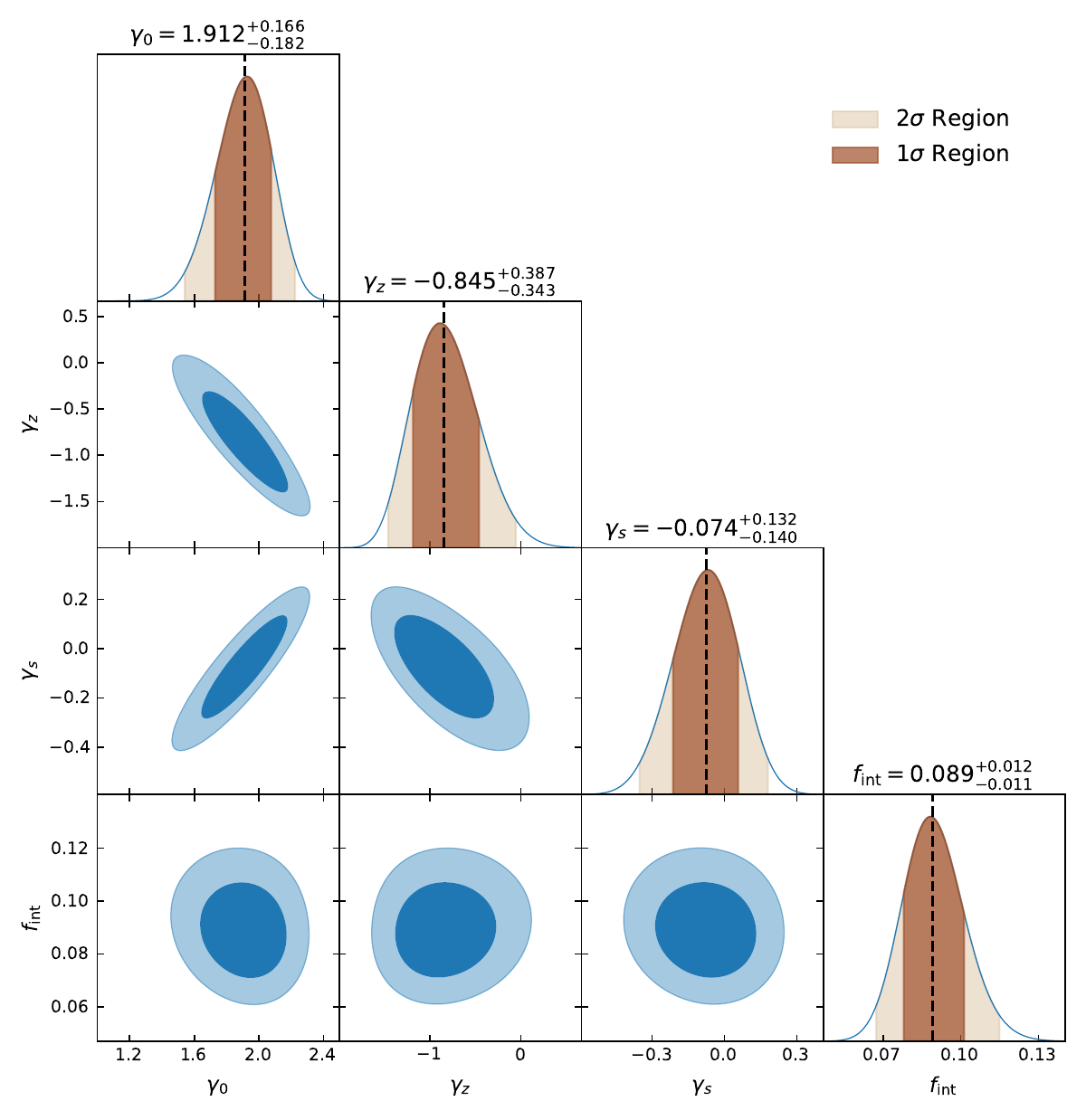}
\includegraphics[width=.49\textwidth]{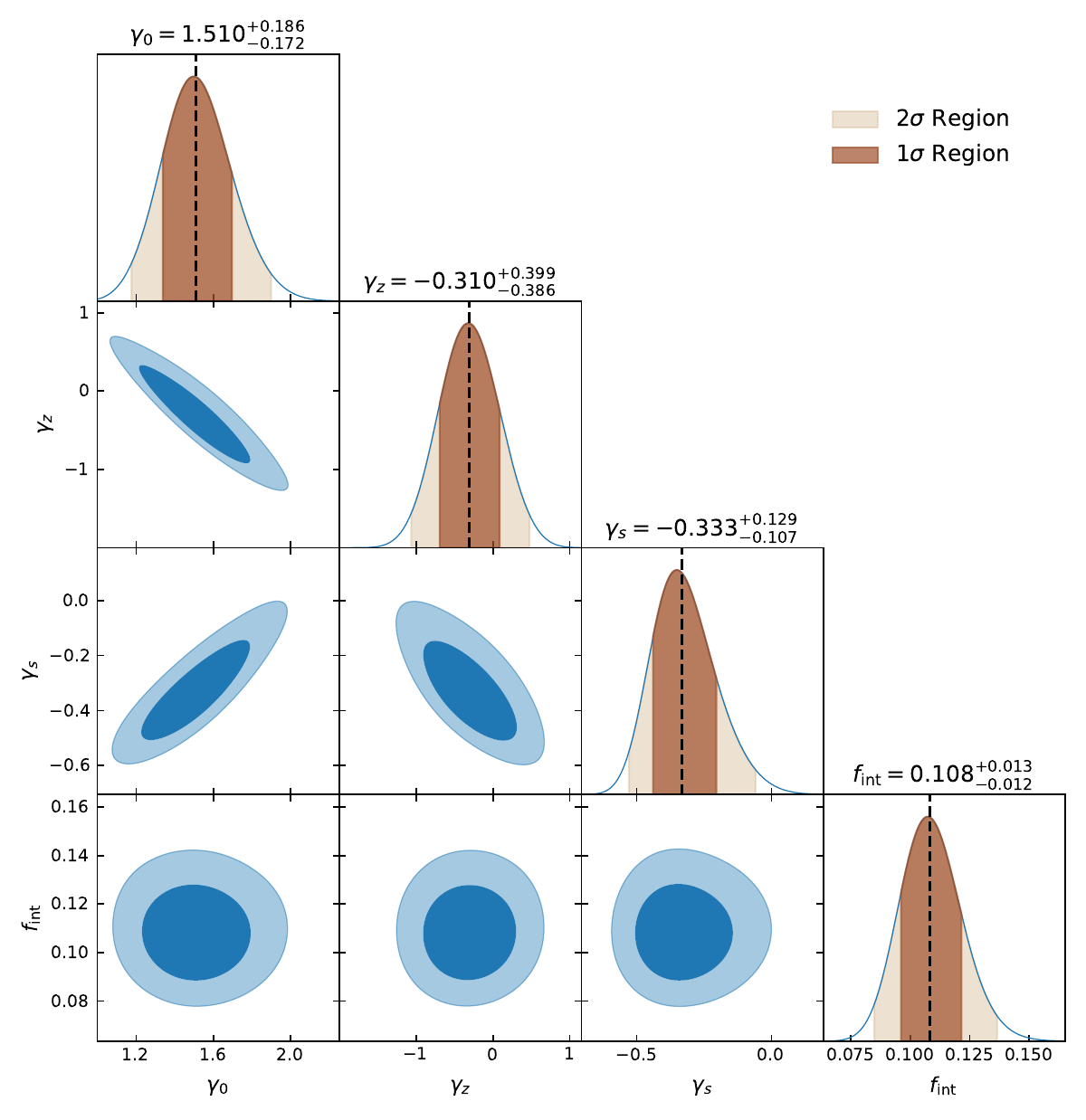}
\caption{P3 individual-slope power-law posteriors with fixed $\beta=0$ (left) and $\beta=0.18$ (right). The structural predictor and other model assumptions are the same as in the free-beta P3 panels.}\label{fig:p3_fixed}\end{figure*}

Figures~\ref{fig:corner_p2_powerlaw_free} and \ref{fig:corner_p2_powerlaw_fixed} show the P2 power-law cases; Figs.~\ref{fig:corner_p2_hernquist_free} and \ref{fig:corner_p2_hernquist_fixed} show Hernquist; Figs.~\ref{fig:corner_p2_sersic_free} and \ref{fig:corner_p2_sersic_fixed} show S\'ersic $n=4$. Figures~\ref{fig:p3_flat}, \ref{fig:p3_gauss}, and \ref{fig:p3_fixed} give the P3 controls. Fixed-beta triangles omit beta because it is not sampled. The plotted $\gamma_0$ remains the zero-redshift intercept, and $f_{\rm int}$ is fractional dispersion scatter. The panels display the parameter correlations for each tracer and anisotropy treatment. The P3 posterior uses the lensing-based predictor defined in the structural-control subsection, a uniform $-1.5<\gamma_s<1.5$ prior, and the same sampling and convergence requirements as P2. It remains a power-law structural control and does not extend the curved-tracer analysis to P3.
\clearpage
\clearpage
\clearpage
\clearpage
\clearpage
\clearpage

\end{document}